\documentclass{iopart}
\pdfoutput=1

\usepackage{graphicx}
\input{epsf}
\def\openone{{\leavevmode\hbox{\small1\kern-3.55pt\normalsize1}}}%
\usepackage{amssymb}
\begin{document}

\title{PageRank of integers} 
\author{K.M. Frahm$^1$, A.D. Chepelianskii$^2$ and D.L. Shepelyansky$^1$}
\address{$^1$Laboratoire de Physique Th\'eorique du CNRS, IRSAMC, 
Universit\'e de Toulouse, UPS, 31062 Toulouse, France}
\address{$^2$ Cavendish Laboratory, Department of Physics, 
University of Cambridge, CB3 0HE, United Kingdom}


\begin{abstract}
We build up a directed network tracing links 
from a given integer to its divisors and analyze the properties of 
the Google matrix of this network. The PageRank vector of this matrix 
is computed numerically and it is shown that its probability is inversely 
proportional to the PageRank index thus being similar to the Zipf law and 
the dependence established for the World Wide Web.
The spectrum of the Google matrix of integers is characterized by a large 
gap and a relatively small number of nonzero eigenvalues.
A simple semi-analytical expression for the PageRank of integers is 
derived that allows to find this vector for matrices of billion size.
This network provides a new PageRank order of integers. 

\end{abstract}


\submitto{\JPA}
\maketitle

\section{Introduction}
The number theory \cite{numbersbook} is the fundamental branch of mathematics
where the theory of prime numbers, besides its beauty, finds
important cryptographic applications \cite{primesbook}.
It is established that the methods of the Random Matrix theory
and quantum chaos find their useful  applications
for the understanding of properties of 
prime numbers and the Riemann zeros \cite{berry,keating,srednicki}. 

In this work we propose another matrix approach to the number theory 
based on the Markov chains \cite{markov} and the 
Google matrix \cite{brin}. The later finds important applications
for the information retrieval and Google search engine of
the World Wide Web (WWW) \cite{meyerbook}. The right eigenvector
of the Google matrix with the largest eigenvalue  
is known as the PageRank vector. 
The elements of this vector
are non-negative and have the meaning of probability to find
a random surfer on the network nodes. 
The PageRank algorithm ranks all websites 
in a decreasing order of  components of the PageRank vector
(see e.g. detailed description at \cite{meyerbook}).
Here, we propose a natural way to construct the Google matrix
of positive integers using their division properties.
We study the statistical properties of the
PageRank vector of this matrix and discuss the properties of a new
order of integers given by this ranking.
The properties of the eigenvalues and eigenvectors
are also discussed.

The paper is constructed as follows: in Section 2
we give the definition of the Google matrix of integers,
the properties of its PageRank vector are analyzed in Section 3,
the analysis of spectral properties is given in Section 4,
the analytical expressions for the PageRank vector
are presented in Sections 4,5
and the discussion of the results is presented in Section 6.  

\section{Google matrix of integers}
The elements of the Google matrix $G(\alpha)$ of 
a directed network with $N$ nodes are given by
\begin{equation}
  G_{mn}(\alpha) = \alpha S_{mn} + (1-\alpha) / N \;\; .
\label{eq1} 
\end{equation} 
Here the matrix $S$ is obtained by normalizing 
to unity all columns of the adjacency matrix $A_{mn}$,
and replacing the elements of columns with only zero elements, 
corresponding to dangling nodes, by $1/N$.
An element $A_{mn}$ of the adjacency matrix is equal to unity
if a node $n$ points to node $m$ and zero otherwise.
The damping parameter $\alpha$ in the WWW context describes the probability 
$(1-\alpha)$ to jump to any node for a random surfer. 
The value $\alpha = 0.85$  gives
a good classification of pages for WWW \cite{meyerbook}.
The matrix $G$ belongs to the class of Perron-Frobenius 
operators \cite{meyerbook}, 
its largest eigenvalue 
is $\lambda = 1$ and the other eigenvalues obey $|\lambda| \le \alpha$. 
In typical WWW networks the eigenvalue $\lambda=1$
is strongly degenerate at $\alpha=1$ (see e.g. \cite{univpr})
and the introduction of $\alpha < 1$ becomes compulsory to define a unique 
right eigenvector at $\lambda=1$ and to 
ensure the convergence
of the PageRank vector by the power iteration method \cite{meyerbook}.
The right eigenvector 
at $\lambda = 1$ gives the probability $P(n)$ to find 
a random surfer at site $n$ and
is called the PageRank. Once the PageRank is found, 
all nodes can be sorted by decreasing probabilities $P(n)$
and an increasing index $K(n)$. 
The node rank is then given by index $K(n)$ which
reflects the  relevance of the node corresponding to a positive integer $n$. 
The PageRank dependence on $K$ is well
described by a power law $P(K) \propto 1/K^{\beta_{in}}$ with
$\beta_{in} \approx 0.9$. This is consistent with the relation
$\beta_{in}=1/(\mu_{in}-1)$ corresponding to the average
proportionality of PageRank probability $P(n)$
to its in-degree distribution $w_{in}(k) \propto 1/k^{\mu_{in}}$
where $k(n)$ is a number of ingoing links for a node $n$  
\cite{meyerbook}. For the WWW it is established that
for the ingoing links $\mu_{in} \approx 2.1$ (with $\beta_{in} \approx 0.9$)
while for out-degree distribution
$w_{out}$ of
outgoing links a power law has the exponent  $\mu_{out} \approx 2.7$
\cite{donato,upfal}. Here we analyze properties of PageRank and use the notation
$\beta=\beta_{in}$.

\begin{figure} 
\begin{indented}\item[]
\begin{center}
\includegraphics[width=0.80\textwidth]{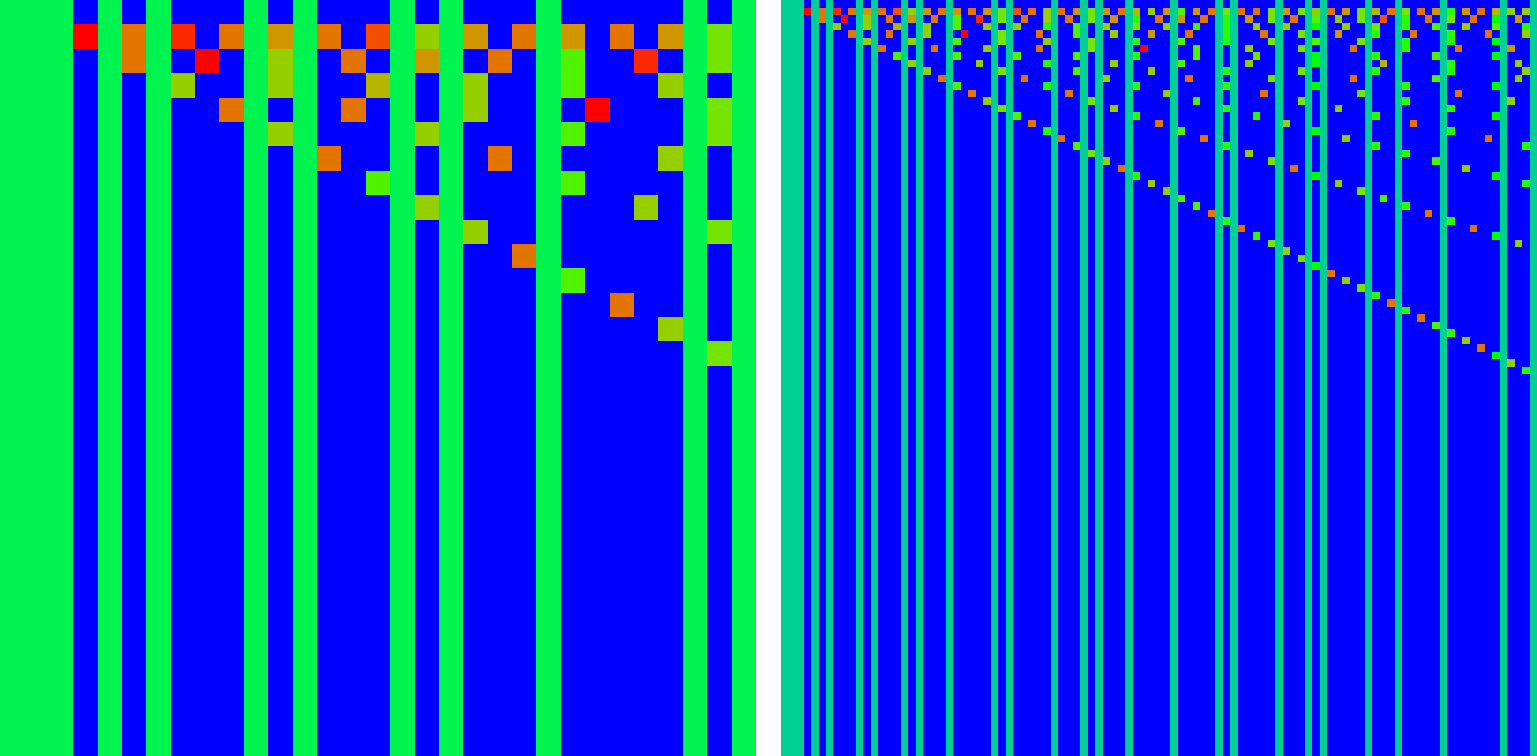}
\end{center}
\vglue -0.2cm
\caption{The Google matrix of integers: the amplitudes of the matrix elements
$G_{mn}$ at $\alpha=1$ are shown by color with blue for minimal zero elements
and red for maximal unity elements, with
$1 \leq n \leq N$ corresponding to $x-$axis (with $n=1$ corresponding 
to the left column) 
and $1 \leq m \leq N$ for 
 $y-$axis (with $m=1$ corresponding to the upper row). 
The matrix sizes are $N=31$ in the left panel and $N=101$ 
in the right panel. 
}
\label{fig1}
\end{indented}
\end{figure} 

To construct the Google matrix of integers, 
we define for $m,n\in\{1,\,\ldots,\,N\}$ the adjacency matrix by 
$A_{mn}=k$ where the $k$ is a ``multiplicity'' defined $k$ as the largest 
integer such that $m^k$ is a divisor of $n$ and if $1<m<n$, 
and $k=0$ if $m=1$ or $m=n$ or if $m$ is not a divisor of $n$. 
Thus we have $k=0$ if $m$ is not 
a divisor of $n$ and $k\ge 1$ if $m$ is a divisor of $n$ different 
from $1$ and $n$. The total size $N$ of the matrix is 
fixed by the maximal considered integer.

This defines a network where an integer number $n$ is linked to its 
divisors $m$ different from $1$ and $n$ itself and where the transition 
probability is proportional to the multiplicity $k$, the number of times 
we can divide $n$ by $m$. The number $1$ and the prime numbers are therefore 
not linked to any other number and correspond to dangling nodes in the 
language of WWW networks. For example, the number $n=24$ has links pointing to
$m(k)=2(3),\, 3(1),\, 4(1),\, 6(1),\, 8(1),\, 12(1)$ 
(multiplicity is given in the brackets)
so that the nonzero matrix elements in this column are
$3/8,\,1/8,\,1/8,\,1/8,\,1/8,\,1/8$ respectively. 
We find the total number of links $N_\ell=\sum_{mn} A_{mn}$, 
taking into account the multiplicity, to be
$N_\ell= 6005$ at $N=1000$, $N_\ell= 1066221$ at $N=10^5$,
$N_\ell= 152720474$ at $N=10^7$, $N_\ell= 19877650264$ at $N=10^9$.
The fit of the dependence
$N_\ell=N\,(a_\ell+b_\ell\ln N$) gives $a_\ell= -0.901 \pm 0.018$,
$b_\ell= 1.003 \pm 0.001$.

From the adjacency matrix $A$ we first construct a matrix $S_0$ 
by normalizing the sum in each column, containing at least 
one non-zero element, to unity and the matrix $S$ is obtained 
from $S_0$ by replacing the elements of columns with only zero elements, 
corresponding to dangling nodes 1 and prime numbers, by $1/N$.
The Google matrix $G$ is finally obtained from $S$ by Eq. (\ref{eq1}) 
for an arbitrary damping factor. The PageRank is the right eigenvector of 
the matrix $G$ with the maximal eigenvalue $\lambda=1$: $GP=\lambda P =P$.

The examples of the Google matrix $G$ at $\alpha=1$ for $N=31,\, 101$
are shown in Fig.~\ref{fig1}. We see that most elements are concentrated 
above the main matrix diagonal
since the divisors $m$ are smaller than the number $n$ itself.
The only exceptions are given by the columns at $1$ and the prime numbers 
$p$ which have no divisors (apart from $1$ and $p$) 
and hence they correspond to the dangling nodes
with no direct links pointing to them. The amplitude of the elements in 
these columns is uniformly $1/N$. The structure of the matrix clearly 
shows the presence of diagonals $m = n/2,\, n/3,\,\ldots$
corresponding to the small divisors $m'=2,\,3,\,\ldots$, 
which appear rather often
in the division of integers. This structure is preserved up to the 
largest size $N=10^9$ considered in this work. 

As we will see in Section \ref{sec_spec}, the eigenvalue $\lambda_0=1$ 
of the matrix $S$ 
is non-degenerate (contrary to typical realistic WWW 
networks \cite{univpr}) and in addition its spectrum has a large gap with
$\lambda_0$ and the other eigenvalues $|\lambda_i| < 0.6$. In such a case
the PageRank vector $P(K)$ has a very small variation when the damping factor 
$\alpha$ is changed in the range $0.85 \leq \alpha \leq 1$ and 
the convergence of the power method to calculate the PageRank is well assured, 
actually quite fast, even for the damping parameter $\alpha=1$. Therefore, we 
limit in this work our studies to the case $\alpha=1$ at 
which $G$ coincides with the matrix $S$ and from now on we denote 
$S$ as ``the Google matrix''. 

\section{PageRank order of integers}

We first determine the PageRank vector of the Google matrix numerically 
by the power iteration method \cite{meyerbook} or by the Arnoldi method 
\cite{arnoldibook} using an Arnoldi dimension of size $n_A$, 
which allows to find several eigenvalues and eigenvectors 
with largest $|\lambda|$ for a full matrix size of a few millions 
(see more details in \cite{univpr,ulamfrahm}).

\begin{figure} 
\begin{indented}\item[]
\begin{center}
\includegraphics[width=0.80\textwidth]{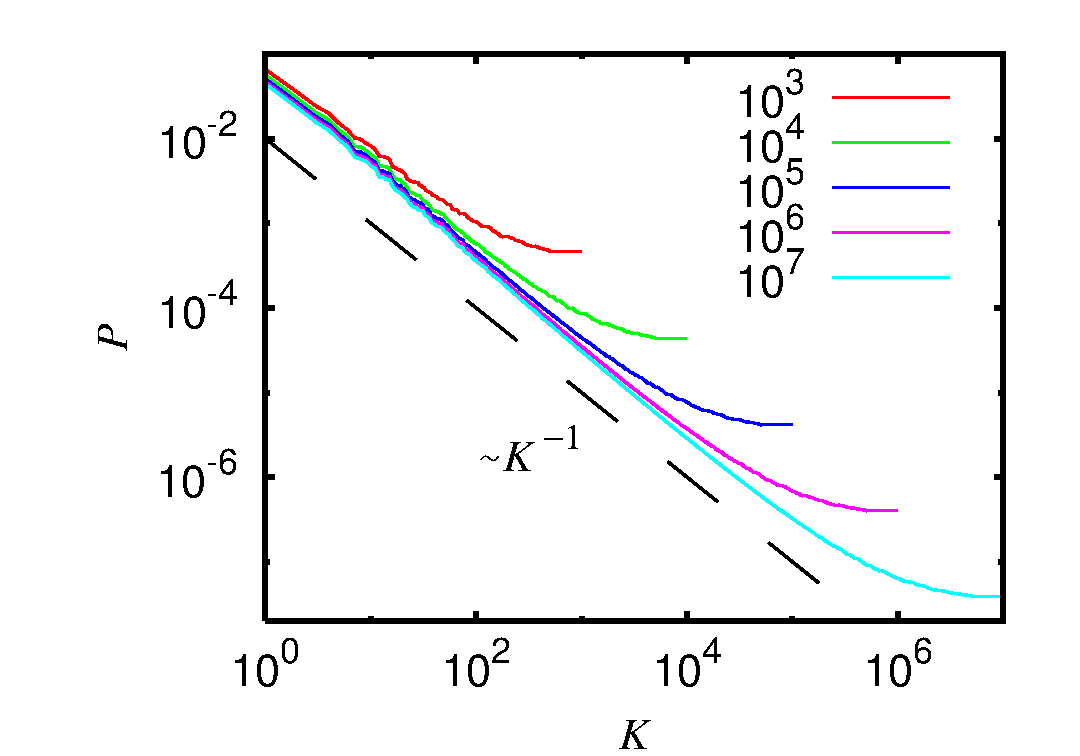}
\end{center}
\vglue -0.2cm
\caption{Dependence of PageRank probability $P(K)$
on PageRank index $K$ for the matrix sizes 
$N=10^3$, $10^4$, $10^5$, $10^6$, $10^7$;
the dashed straight line shows the Zipf law dependence $P \sim 1/K$.
}
\label{fig2}
\end{indented}
\end{figure} 

\begin{figure} 
\begin{indented}\item[]
\begin{center}
\includegraphics[width=0.80\textwidth]{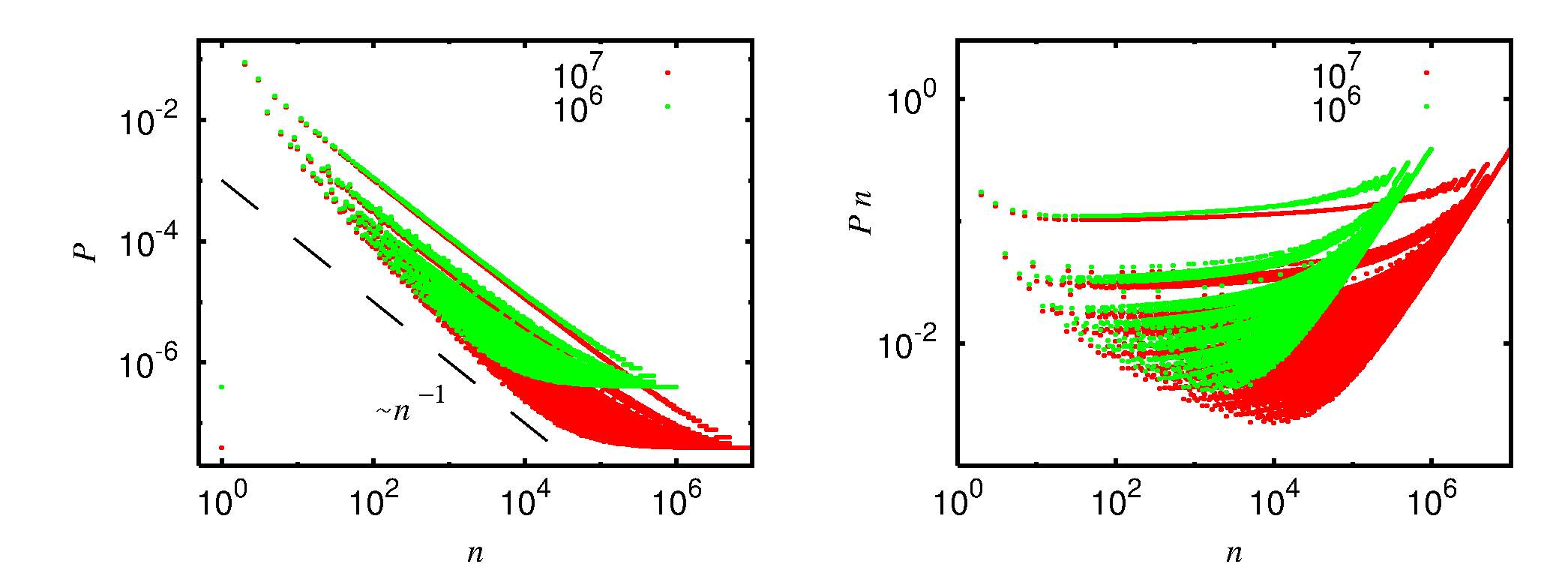}
\end{center}
\vglue -0.2cm
\caption{Dependence of PageRank probability $P$ on the integer number $n$  
for matrix sizes $N=10^6, 10^7$ (left panel green and red points respectively),
and rescaled probability $n P$ on $n$ (right panel); data are 
shown in log-log scale.}
\label{fig3}
\end{indented}
\end{figure}

The dependence of PageRank probability $P(K)$ on PageRank index $K$
is shown in Fig.~\ref{fig2}. We see that with the growth 
of the system size $N$
the dependence $P(K)$ converges to a fixed distribution $P(K)$
on initial $K \leq N/10$ values 
with  the tail of distribution $P(K)$ at $K >N/10$
which is sensitive to the cut-off at the finite matrix size $N$.
In the convergent part a formal fit (for $10<K<10^5$)
gives the dependence $P \sim A/K^{\beta}$ with 
$\ln A = 0.0431 \pm 0.00049$,
$\beta = 1.040 \pm 0.0015$ 
being close to the Zipf law with $\beta=1$ \cite{zipf}.
The small value of $\beta -1$ indicates that there can 
be a logarithmic correction. Indeed, the fit 
$1/(P K) = a_1+b_1 \ln K$ (for $10<K<10^3$) gives the values 
$a_1=16.050 \pm 0.187$, $b_1=2.468 \pm 0.036$.
Thus, it is possible that in the limit of $N \rightarrow \infty$
we have the asymptotic behavior $P \sim 1/(K \ln K)$.
Such a scaling looks to be more probable due to usual
logarithmic corrections in the density of primes \cite{primesbook}.
However, for the available finite matrix sizes the regime 
of linear bevahior of $1/(PK)$ versus $\ln K$ is quite limited and 
it is not obvious to distinguish between the above two 
fitting dependencies.
\begin{figure} 
\begin{indented}\item[]
\begin{center}
\includegraphics[width=0.80\textwidth]{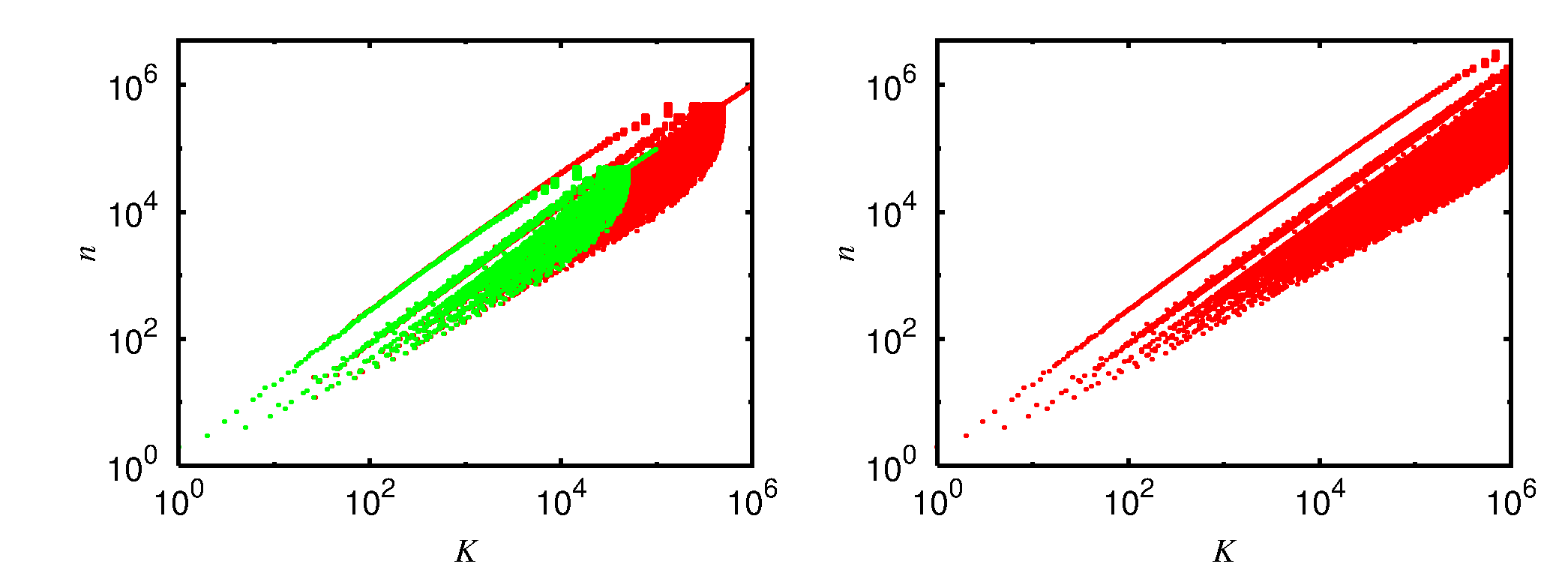}
\end{center}
\vglue -0.2cm
\caption{Dependence of the integer number $n$ on the 
PageRank index $K$ for 
sizes $N=10^5, 10^6$ (left panel green and red points respectively),
and $10^7$ (right panel); data are shown in log-log scale.
}
\label{fig4}
\end{indented}
\end{figure}

The dependence of PageRank probability $P$ on integer 
index $n$ is shown in Fig.~\ref{fig3}.
It is characterized by a global decay $P \propto 1/n$
with the presence of various branches which are especially well
visible for the rescaled quantity $n P$. This structure 
is preserved with the increase of matrix size for the values of $n < N/100$.
The direct check shows that the highest plateau 
corresponds to the prime numbers $p$.
\begin{figure} 
\begin{indented}\item[]
\begin{center}
\includegraphics[width=0.80\textwidth]{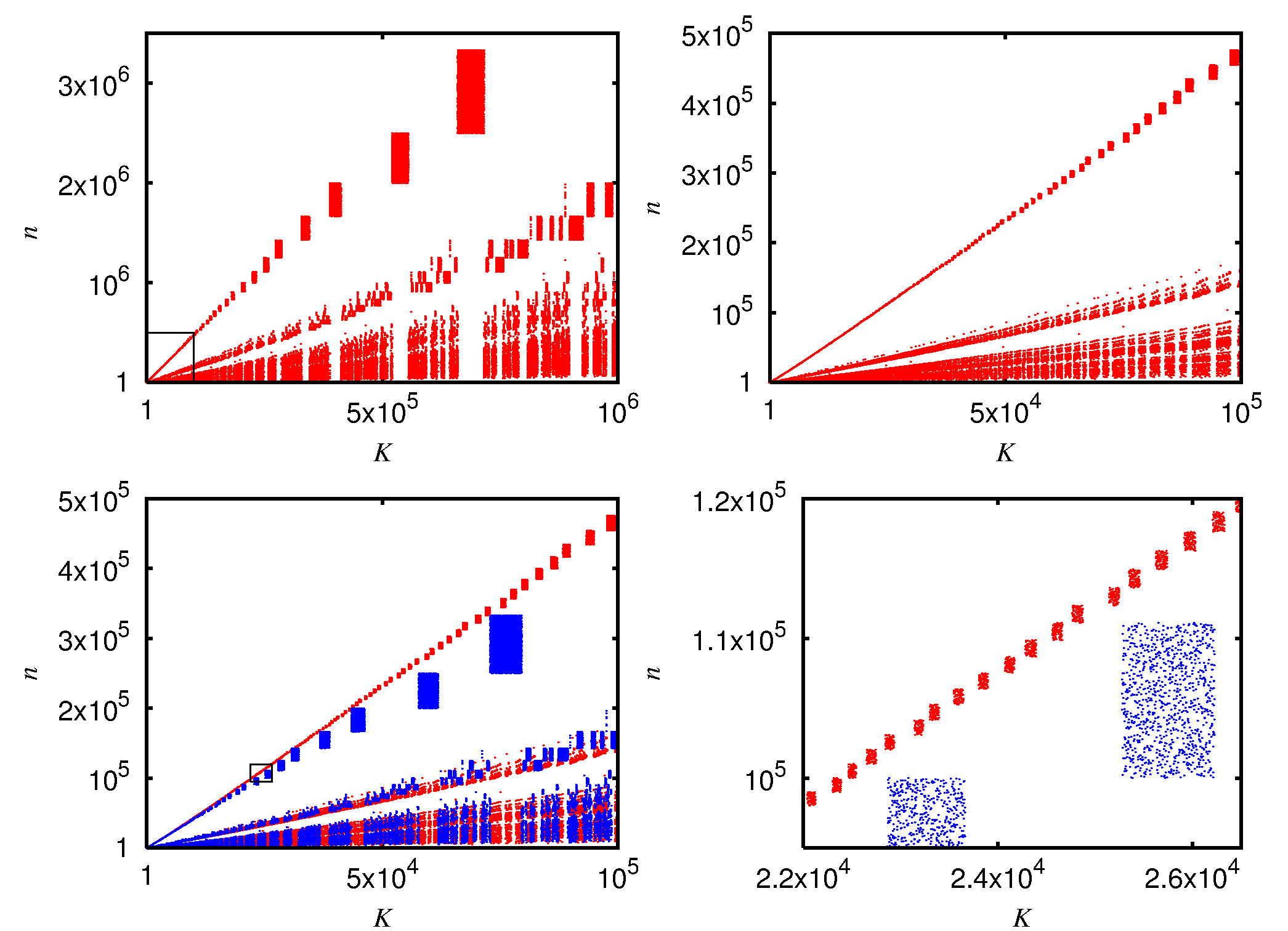}
\end{center}
\vglue -0.2cm
\caption{Top panels: dependence of the integer number $n$ on
PageRank index $K$ for 
size $N=10^7$ shown by red points (left panel), 
right panel shows zoom of data in a rectangle
from left panel.
Bottom panels: in addition to data of top right panel data for $N=10^6$
are shown (left panel), right panel shows zoom of data in a rectangular
region from left panel. Data are shown in usual scale.
}
\label{fig5}
\end{indented}
\end{figure}

Another way to analyze the structures visible in Fig.~\ref{fig3}
is to consider the dependence of $n$ on the PageRank index $K$
obtained from the PageRank probability $P(K_n)$.
In fact $K$ gives a new order of integers
imposed by the PageRank. 
The dependence $n(K)$ is shown in Fig.~\ref{fig4}
on a large scale. In a first approximation we find the layered structure 
with a sequence of parallel lines $n \propto K$.
This global structure is preserved with the increase
of the matrix size from $N=10^5$ to $10^7$.

A more detailed view of this structure is shown in Fig.~\ref{fig5}.
There are well defined separated branches with approximately 
linear dependence $n \approx \kappa K$ with
$\kappa \approx 4.5$ for the highest branch which 
corresponds to the highest plateau in Fig.~\ref{fig3} (right panel).
This branch contains only primes. The lower branch
contains semi-primes (product of two primes) and so on down to smaller
an smaller values of $\kappa$. The whole structure looks to have 
a self-similar structure as it shows a zoom to a smaller scale.
The increase of the size $N$ gives some modifications of the 
structure keeping its global pattern (see Fig.~\ref{fig5} bottom panels).
There is a certain clustering on the $(n,K)$ plane of rectangles 
containing close values of $K$ and integer numbers $n$. The rectanglers 
in the upper prime-branch contain exclusively prime numbers for $n=p$. 
Note that the neighboring non-prime values appear 
in other rectanglers on the right side for larger values of $K$. 
For example, in the bottom left panel of Fig.~\ref{fig5} 
we have a rectangle
at $K\sim 2.6\times 10^4$ and $n\sim 10^5$ with primes 
but there is at $K\sim 7\times 10^4$ 
another rectangle of semi-primes, also with values $n\sim 10^5$. 

The direct analysis shows that the rectangles correspond to 
flat plateaux with degenerate values of $P(K_n)$ which appear for finite 
matrix size $N$. This degeneracy results 
from only rational numbers appearing in the elements of the Google matrix
and from its very sparse structure. 
Inside such flat regions the ordering in $K$
is somewhat arbitrary and depends on the precise sorting algorithm used. 
The $K$ index shown in Fig.~\ref{fig5} was obtained by the Shellsort method 
that may indeed produce a quite random ordering for degenerate values 
thus generating the rectanglers seen in Fig.~\ref{fig5}. 
We have verified that 
when using a modified sorting algorithm with a secondary criterium, to sort 
with increasing $n$ inside a degenerate region, the rectangles 
are replaced by lines from the left bottom corner to the right top corner.
With increasing values of $N$ these rectangles are reduced in size. 
We numerically find that the first degenerate plateau 
appears at $K=K_d$ and that this number increases with the matrix size $N$,
e.g. $K_d=27$ at $N=1000$, $177$ at $10^5$,
$1287$ at $10^7$, $10386$ at $10^9$.
This dependence is well described by the fit
$K_d=a_d K^{b_d}$ with $a_d=1.284 \pm 0.078$, $b_d=0.432 \pm 0.004$.
We return to the discussion of 
the convergence at large $N$ a bit later.
\begin{figure} 
\begin{indented}\item[]
\begin{center}
\includegraphics[width=0.80\textwidth]{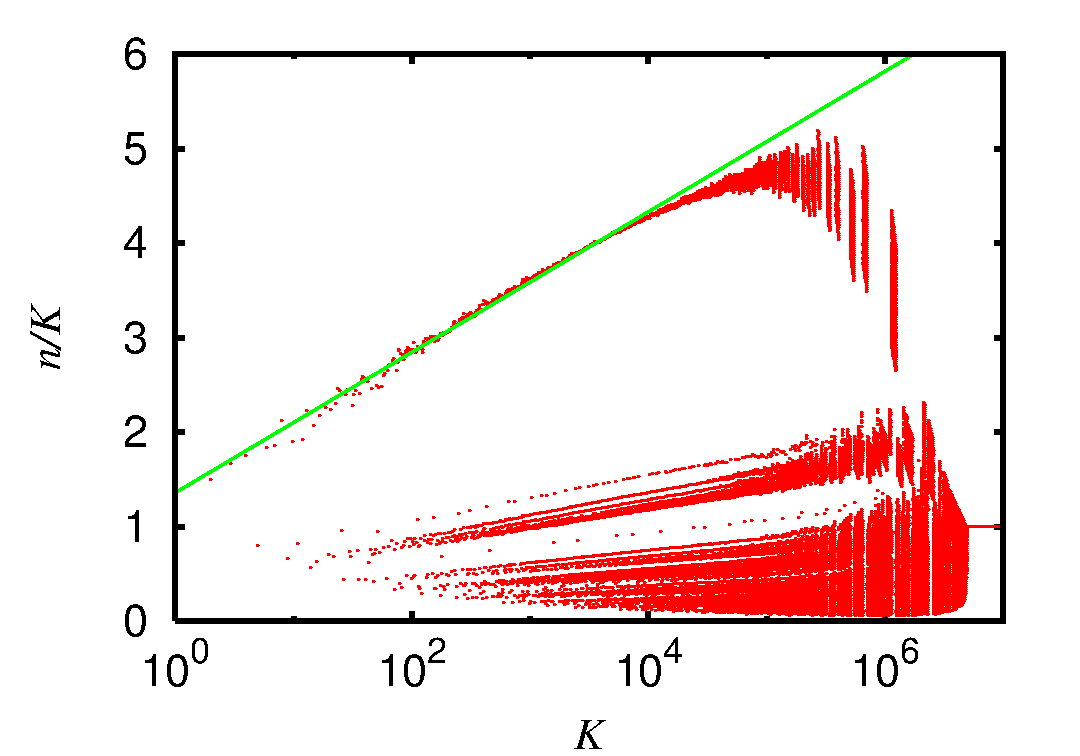}
\end{center}
\vglue -0.2cm
\caption{Dependence of the ratio  $n/K$ on
the PageRank index $K$ for 
size $N=10^7$; data are shown in semi-log scale.
The straight line shows the fit dependence
$n/K=a_2+ b_2\ln K$ for the upper branch
in the range $10 \leq K \leq 10^4$
with $a_2=1.3583 \pm 0.0099$, $b_2=0.3227 \pm 0.0014$.
}
\label{fig6}
\end{indented}
\end{figure}

Since we find an approximate linear growth of $n$ with $K$
inside each branch it is useful to consider the
dependence of the ratio $n/K$ on $K$ which is 
shown in Fig.~\ref{fig6}. The upper branch of primes is 
well described by the dependence 
$n/K = 0.322 \ln K +1.358$ that shows that in the previous
relation $\kappa$ is not a constant but grows
logarithmically with $K$. 
We have an approximate relation $b_2 =0.322 \approx 1/b_1 =1/2.468$.
The lower branches also have an approximately logarithmic growth of
the ratio $n/K$ with $K$.
\begin{figure} 
\begin{indented}\item[]
\begin{center}
\includegraphics[width=0.80\textwidth]{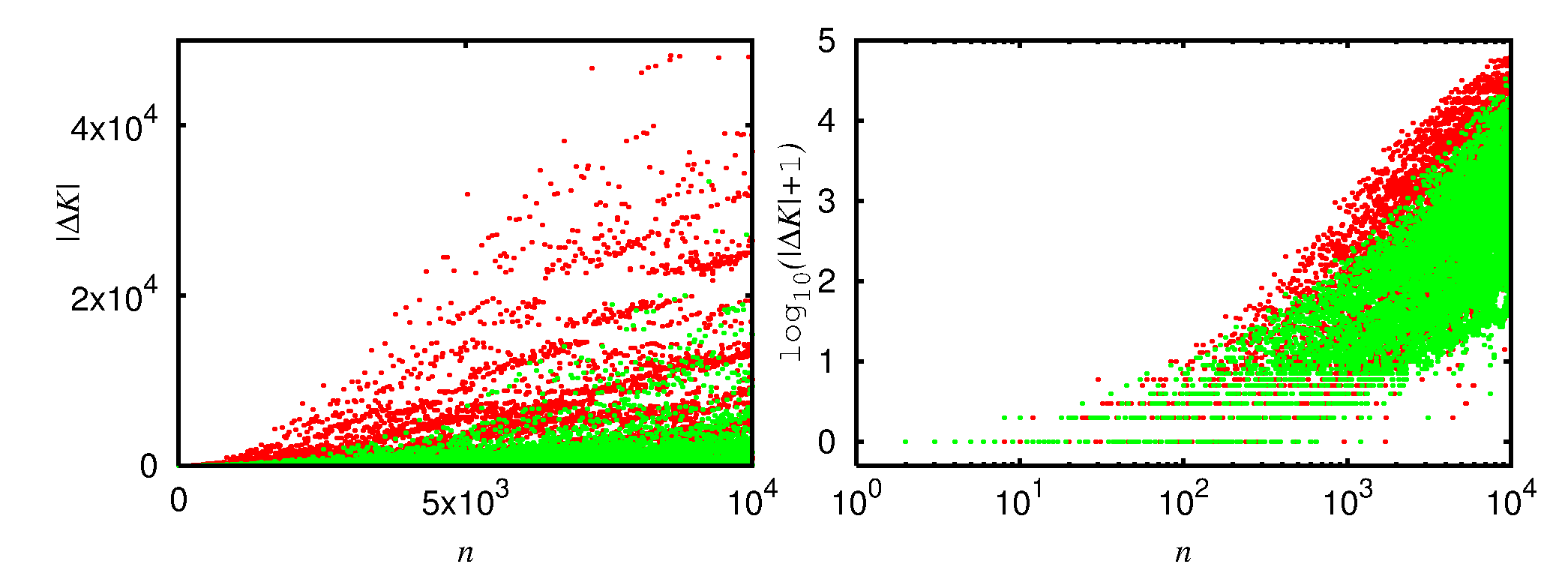}
\end{center}
\vglue -0.2cm
\caption{Dependence of $|\Delta K| = |K_n(N_2)-K_n(N_1)|$
on the integer $n$ for matrix sizes $N_1=10^6, N_2=10^7$
(green points) and
$N_1=10^5, N_2=10^6$ (red points).
Left and right panels show the same data either 
in normal or in log-log scales.}
\label{fig7}
\end{indented}
\end{figure}

Finally, let us discuss the stability of the PageRank order
of integers in respect to the variation of the matrix size $N$.
The dependence $P(K)$ is definitely converging to a fixed
function for $K \ll N$ as it is well seen in Fig.~\ref{fig2}.
However, for a fixed integer $n$ its PageRank index $K_n$
has a visible variation with the increase of matrix size $N$.
These variations are visible in Fig.~\ref{fig5} (bottom panels).
At the same time the global structure of the $K_n$ or $n(K)$
dependence shows signs of convergence with the growth of $N$.
A more detailed analysis of variation of $\Delta K = |K_n(N_1)-K_n(N_2)|$
for two matrix sizes $N_2=10N_1$ is shown in Fig.~\ref{fig7}.
We see that there is a significant decrease of variations
$\Delta K$ with increase of $N_1$, even if a small
changes of $K_n$ values are visible even at relatively low $n \sim 100$.
On the basis of these data we make a conjecture
that in the limit of $N \rightarrow \infty$
we will have a convergence to a fixed PageRank order
of integers $K_n$. However, we expect that this convergence is very slow,
probably logarithmic in $N$, thus being the reason that even at 
$N=10^7$ we find some variations in $K_n$.
We note that the density of states of Riemann zeros
also shows very slow convergence so that
enormously large values of $n \sim N \sim 10^{20}$
are required to obtain stable results \cite{berry,keating}.


\section{Spectral properties of the Google matrix of  integers}
\label{sec_spec}

\subsection{Arnoldi method}

To study numerically the spectrum of the Google matrix $S=G$ of integers 
at $\alpha=1$
we first employ the Arnoldi method \cite{arnoldibook,ulamfrahm}. 
This method uses a normalized initial vector $\xi_0$ and 
generates a {\em Krylov space} 
by the vectors $S^j\,\xi_0$ for $j=0,\,\ldots,\,n_A-1$ where 
$n_A$ is called the Arnoldi dimension. Using Gram-Schmidt orthogonalization 
one determines an orthogonal basis of the Krylov space and the matrix 
representation of $S$ in this basis. This provides a matrix $\bar S$ of modest 
dimension $n_A$ of Hessenberg form which can be diagonalized by standard 
QR-methods and whose eigenvalues, called {\em Ritz eigenvalues}, are 
in general very accurate approximations of the largest eigenvalues of the 
original (very large) matrix $S$. 

\begin{figure} 
\begin{indented}\item[]
\begin{center}
\includegraphics[width=0.80\textwidth]{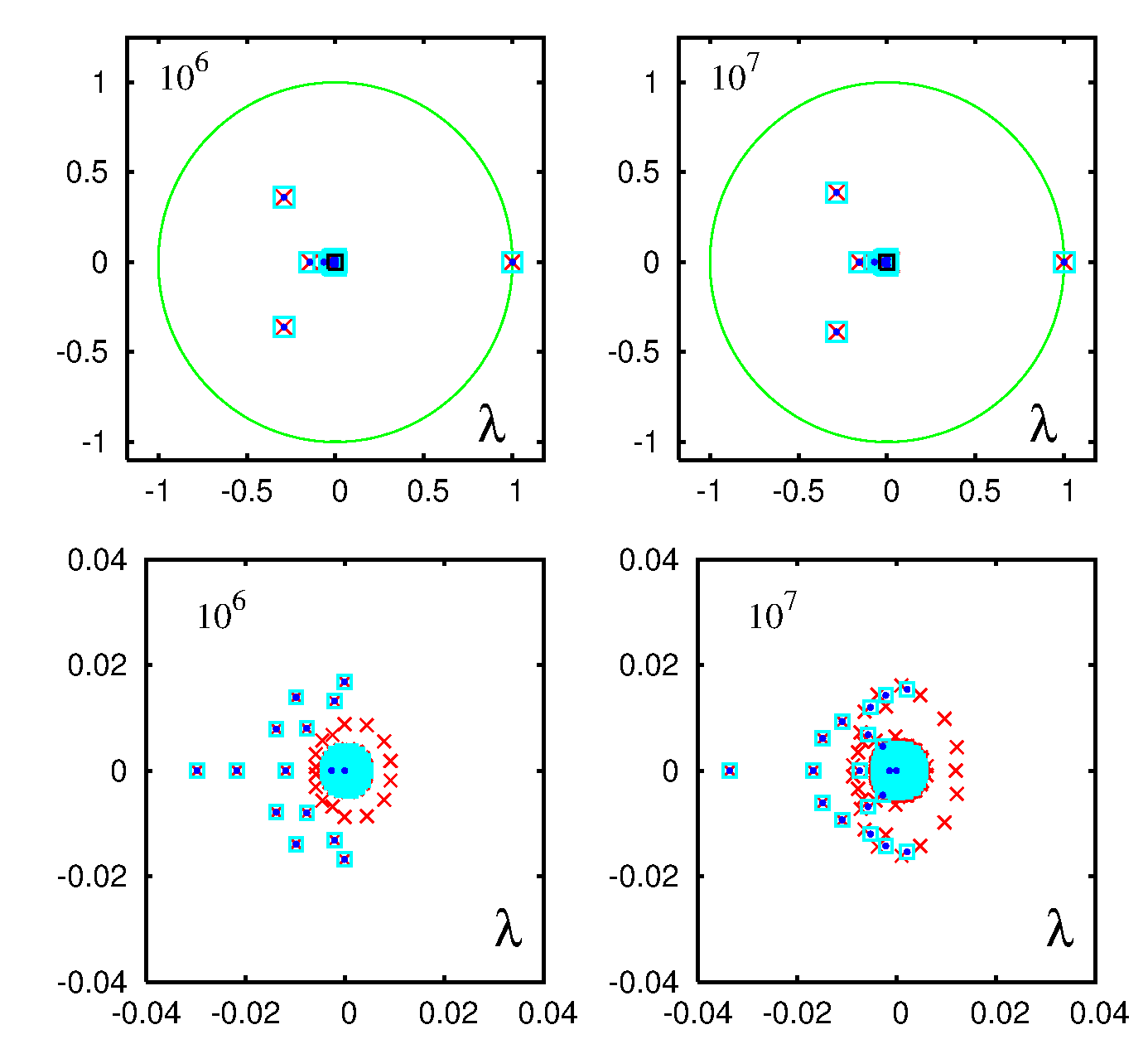}
\end{center}
\vglue -0.2cm
\caption{Spectrum of the Google matrix of  integers
for the matrix size $N=10^6$ (left panels)
and $10^7$ (right panels); the red crosses (light blue squares) are numerical 
data from the Arnoldi method with Arnoldi dimension $n_A=1000$ and 
a random initial vector (with the unit initial vector) and the dark 
blue points are the exact eigenvalues obtained as the zeros of the reduced 
polynomial of Eq. (\ref{eq5}). 
Top panels show the whole spectrum and the 
bottom panels show a zoom of the region represented by black squares
in the top panels. The eigenvalues 
have significantly higher accuracy for the Arnoldi method
with unit initial vector. The unit circle $|\lambda|=1$ is
shown in green.
}
\label{fig8}
\end{indented}
\end{figure}

In this work we have used the Arnoldi dimension $n_A=1000$ and two different 
initial vectors, first a random initial vector and second a uniform initial 
vector with identical components $1/\sqrt{N}$ (thus normalized by the 
Euclidean norm $\|(\cdots)\|_2$). 
The spectrum of the matrix $S$ is shown in Fig.~\ref{fig8}
for two sizes $N=10^6,\,10^7$. We see that there are only three eigenvalues 
within the ring $0.05 <|\lambda| < 0.5$ while the majority of 
eigenvalues is concentrated inside a range of $|\lambda| < 0.05$. 
The first few largest eigenvalues are accurately obtained from both initial 
vectors used for the Arnoldi method and also coincide (up to numerical 
precision) with the eigenvalues determined by a 
semi-analytical approach (see below). 
However for the range $|\lambda| < 0.05$ the situation 
becomes more subtle as it is discussed below. 

We note that Fig.~\ref{fig8} shows a large gap between $\lambda_0=1$ and the next 
eigenvalue thus justifying our above choice of the 
damping factor $\alpha=1$ .

\subsection{Analytical discussion of spectrum}

The Google matrix $S$ at $\alpha=1$ has a very particular 
structure which allows to establish some important properties for the spectrum 
and its eigenvalues. We can write 
\begin{equation}
\label{eq2}
S=S_0+v\,d^{\,T}
\end{equation}
where $v$ and $d$ are two vectors of size $N$ with components $v_n=1/N$ and 
$d_n=1$ for prime numbers $n=p$ or $n=1$ and $d_n=0$ for the other 
non-prime numbers (different from 1). For later use we also introduce 
the vector $e$ with components $e_n=1$ and therefore $v=e/N$. 
In addition $d^{\,T}$ denotes the transposed line vector of $d$. 
The matrix $S_0$ is the contribution that arises from the adjacency matrix 
$A$ by normalizing the non-vanishing columns of the latter and the 
tensor product $v\,d^{\,T}$ represents the values $1/N$ which are put in 
the zero columns of $S_0$ when constructing the full matrix $S$. 
The normalization condition of the non-vanishing columns of $S_0$ can 
formally be written as $e^{\,T}\,S_0=e^{\,T}-d^{\,T}$ which is just the 
line vector with components $0$ for vanishing columns of $S_0$ (for 
prime numbers $n$ or $n=1$) and $1$ non-vanishing columns of $S_0$ 
(for the other non-prime numbers different from $1$). 
This expression provides the useful identity~:
\begin{equation}
\label{eq_useful}
d^{\,T}=e^{\,T}(\openone-S_0)\ .
\end{equation}

Furthermore we observe that the matrix $S_0$ has a triagonal form with 
vanishing entries 
on the diagonals because $(S_0)_{mn}\neq 0$ only if $m$ is a divisor 
of $n$ different from $1$ and $n$ and therefore for any non-vanishing 
matrix element $(S_0)_{mn}$ we have $m\le n/2<n$. This matrix structure 
can also be seen in Fig.~\ref{fig1}. As 
a consequence $S_0$ is nilpotent with $S_0^l=0$ for some integer $l$. In 
the following let us assume that $l$ is the minimal number such that 
$S_0^l=0$. Obviously in our model $l=[\log_2(N)]$ is actually a 
very modest number as compared to the full matrix size $N$. 

We now discuss how the form of Eq. (\ref{eq2}) affects the eigenvalues 
of the full matrix $S$. Let $\psi$ be a right eigenvector of $S$ and 
$\lambda$ its eigenvalue~:
\begin{equation}
\label{eq3}
\lambda\psi=S\psi=S_0\psi + C\,v\quad,\quad C=d^{\,T}\,\psi=
\sum_{n\ {\rm prime\ or}\ n=1}^N \psi_n\quad.
\end{equation}
If $C=0$ we find that $\psi$ is an eigenvector of $S_0$. Then $\lambda=0$ 
since the matrix $S_0$ is nilpotent and cannot have non-vanishing eigenvalues. 
The matrix $S_0$ is actually non-diagonalisable and can only be 
transformed to a Jordan form with quite large Jordan blocks and $0$ as 
diagonal element of each of the Jordan blocks. 

Suppose now that $C\neq 0$ implying that $\lambda\neq 0$ since the 
equation $S_0\psi = -C\,v$ does not have a solution for $\psi$ because 
$S_0$ has many zero rows and $v_n=1/N\neq 0$ for each $n=1,\ldots,N$. 
Since $\lambda\neq 0$ the triagonal matrix $\lambda\openone-S_0$ is invertible 
and from Eq. (\ref{eq3}) we obtain~:
\begin{equation}
\label{eq4}
\psi=C\,(\lambda\openone-S_0)^{-1}\,v=
\frac{C}{\lambda}\,\sum_{j=0}^{l-1}\left(\frac{S_0}{\lambda}\right)^j\,v\quad.
\end{equation}
Note that the sum is finite since $S_0^l=0$. The eigenvalue $\lambda$ 
is determined by the condition that this expression of $\psi$ has 
to satisfy the condition $C=d^{\,T}\,\psi$. Multiplying this condition by $\lambda^l/C$ 
we find that $\lambda$ is a zero of the following {\em reduced polynomial} 
of degree $l$:
\begin{equation}
\label{eq5}
{\cal P}_r(\lambda)=\lambda^l-\sum_{j=0}^{l-1}\lambda^{l-1-j}\,c_j=0
\quad,\quad c_j=d^{\,T}\,S_0^j\,v\quad.
\end{equation}
This calculation shows that there are at most $l$ eigenvalues 
$\lambda\neq 0$ of $S$ given as the zeros of this reduced polynomial. 

We note that using $S_0^l=0$ and 
the identity (\ref{eq_useful}) one finds that the coefficients $c_j$ 
obey the following sum rule~:
\begin{equation}
\label{eq_sum}
\sum_{j=0}^{l-1}c_j=d^{\,T}\left(\sum_{j=0}^{l-1}S_0^j\right)\,v
=e^{\,T}(\openone-S_0)(\openone-S_0)^{-1}\,v=1
\end{equation}
since $e^{\,T}\,v=\sum_n v_n=1$. This sum rule ensures that 
$\lambda=1$ is a zero of the reduced polynomial and the 
PageRank as the eigenvector of $\lambda=1$ is obtained from 
(\ref{eq4})~:
\begin{equation}
\label{eq_PR1}
P=C\sum_{j=0}^{l-1}S_0^j\,v\quad,\quad 
C^{-1}=\sum_{j=0}^{l-1}e^{\,T}\,S_0^j\,v
\end{equation}
where the identity for $C^{-1}$ is due to the normalization of $P$. 

Since the degree $l=[\log_2(N)]$ of the reduced polynomial is very modest:
$9\le l\le 29$ for $10^3\le N\le 10^9$, we have determined numerically 
the coefficients $c_j$ which only requires a finite number of successive 
multiplications of $S_0$ to the initial vector $v$ and determined 
the zeros of the reduced polynomial by the very efficient Newton-Maehly 
method in the complex plane. The resulting $l$ eigenvalues (and the trivial 
highly degenerate eigenvalue $\lambda=0$ of $S$) obtained from this 
semi-analytical method are also shown in Fig.~\ref{fig8}. 

The numerical determination of the zeros shows that they are all simple zeros 
of the reduced polynomial but at this point we are not yet sure that 
they are also non-degenerate as far as the full matrix $S$ is concerned. 
In theory we might still have principal vectors $\phi$
associated to some eigenvalue $\lambda\neq 0$ such that 
$S\phi=\lambda\phi+\psi$ with $\psi$ being the eigenvector at 
$\lambda$. 
However, we can exclude this scenario by determining the full 
characteristic polynomial of $S$:
\begin{eqnarray}
\nonumber
{\cal P}_S(\lambda)&=&\det(\lambda\openone-S_0-v\,d^{\,T})\\
\nonumber
&=&\lambda^N\det(\openone -S_0/\lambda)\,
\det\left[\openone-(\openone-S_0/\lambda)^{-1}\,v\,d^{\,T}/
\lambda\right]\\
&=&\lambda^N\left[1-d^{\,T}(\openone-S_0/\lambda)^{-1}\,v/\lambda\right]
=\lambda^{N-l}\,{\cal P}_r(\lambda)
\label{eq6}
\end{eqnarray}
since $\det(\openone -S_0/\lambda)=1$, 
$\det(\openone-u\,w^{\,T})=(1-w^{\,T}\,u)$ 
for arbitrary vectors $u$ and $w$, and the matrix inverse 
has been expanded in a finite sum in a similar way as in Eq. (\ref{eq4}). 
According to Eq. (\ref{eq6}) we observe that the simple zeros of 
${\cal P}_r(\lambda)$ are also 
simple zeros of ${\cal P}_S(\lambda)$ and have therefore an algebraic 
multiplicity equal to one. This proves that there are no principal vectors 
and no non-trivial Jordan-Block structure for $\lambda\neq 0$. 
On the other hand the eigenvalue $\lambda=0$ has algebraic 
multiplicity $N-l$ with many large Jordan-Blocks. 

The $l$-dimensional subspace associated to the eigenvalues $\lambda\neq 0$ 
is according to Eq. (\ref{eq4}) generated by the $l$ vectors 
$v^{(j)}=S_0^j\,v$ with $j=0,\,\ldots,\,l-1$ which form a basis 
of this subspace. Using Eqs. (\ref{eq2}) and (\ref{eq5}),
we may easily determine the matrix representation of $S$ 
with respect to this basis by:
\begin{equation}
\label{eq_represent1}
S\,v^{(j)}=c_j\,v^{(0)}+v^{(j+1)}=\sum_{k=0}^l \bar S_{k+1,j+1}\,v^{(k)}
\quad,\quad j=0,\,\ldots,\,l-1
\end{equation}
where for simplicity of notation for the case $j=l-1$ 
we write $v^{(l)}=0$. The 
$l\times l$-matrix $\bar S$ has the explicit form~:
\begin{equation}
\label{eq_represent2}
\bar S=
\left(\begin{array}{ccccc}
c_0 & c_1 & \cdots & c_{l-2} & c_{l-1} \\
1   & 0   & \cdots & 0 & 0 \\
0   & 1   & \cdots & 0 & 0 \\
\vdots & \vdots & \ddots & \vdots & \vdots \\
0   & 0   & \cdots & 1 & 0 \\
\end{array}\right)\quad.
\end{equation}
One easily verifies that the characteristic polynomial 
${\cal P}_{\bar S}(\lambda)$ of this matrix coincides with the reduced 
polynomial (\ref{eq5}) and its $l$ eigenvalues are therefore exactly the $l$ 
non-vanishing eigenvalues of the full matrix $S$. Using the sum rule 
(\ref{eq_sum}) one notes that the $l$-dimensional 
vector $(1,\,\ldots,\,1)^{\,T}$ is a right eigenvector of $\bar S$ 
with eigenvalue $\lambda=1$ thus confirming the PageRank expression 
$P\propto\sum_{j=0}^{l-1} v^{(j)}$ [see also Eq. (\ref{eq_PR1})]. 

A direct numerical diagonalisation of the matrix (\ref{eq_represent2}) 
is tricky and fails to produce the smaller eigenvalues (below $10^{-2}$) 
due to numerical rounding errors since the coefficients $c_j$ decay very 
rapidly, e.~g. $c_{22}\sim 10^{-38}$ for $N=10^7$ with $l=23$. However, we 
may numerically diagonalize the ``equilibrated'' matrix~: 
$\rho^{-1}\,\bar S\,\rho$ which has the same eigenvalues as $\bar S$ and 
where $\rho$ is a diagonal matrix with diagonal matrix elements 
$\rho_{jj}=1/c_{j-1}$. The eigenvalues obtained from the equilibrated 
matrix coincide very 
precisely (up to numerical precision $10^{-14}$) with the zeros obtained 
from the reduced polynomial by the Newton-Maehly method. 
In Fig.~\ref{fig8}, we also show these $l$ zeros for $N=10^6$ and 
$N=10^7$. Apparently, both variants of the Arnoldi method fail to confirm the 
analytical result that there are only $l$ non-vanishing eigenvalues, a point 
we attribute to the numerical instability of the highly degenerate and 
defective eigenvalue $\lambda=0$ and which we will discuss below.

\begin{figure} 
\begin{indented}\item[]
\begin{center}
\includegraphics[width=0.80\textwidth]{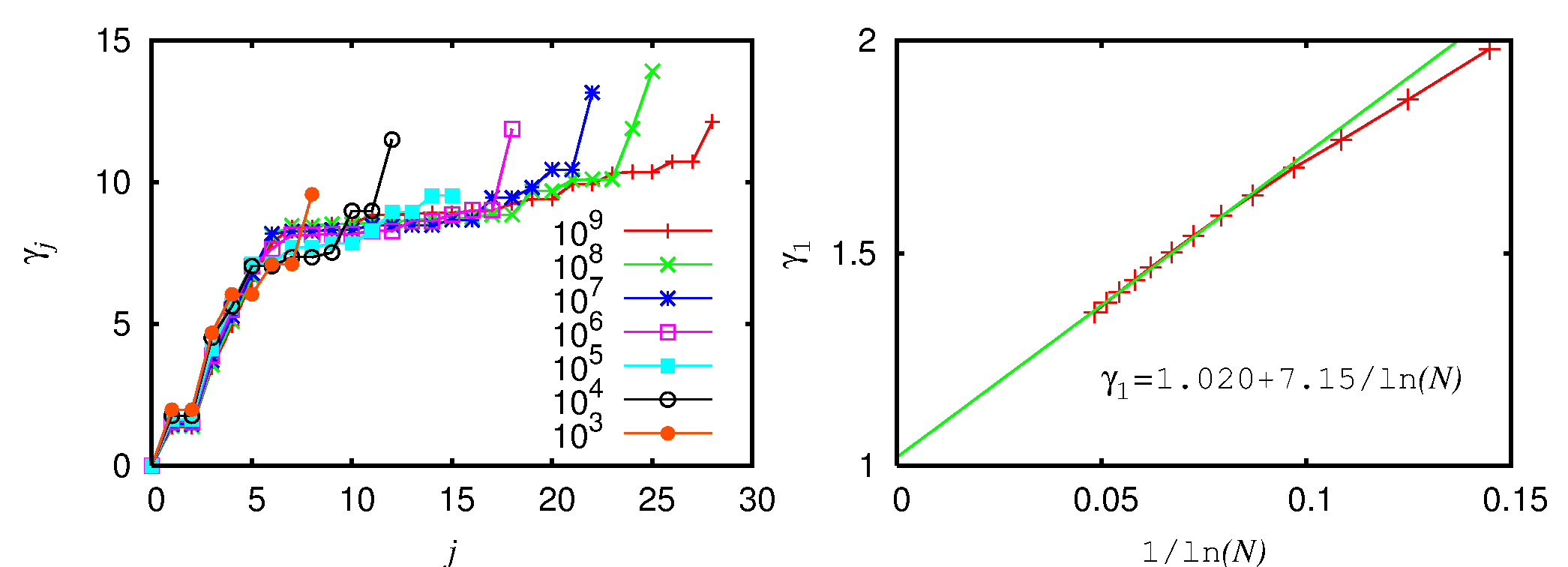}
\end{center}
\vglue -0.2cm
\caption{Left panel: dependence of $\gamma_j=2 \ln|\lambda_j|$ on the 
index $j$ for the $l$ non-vanishing eigenvalues of $S$ and various 
matrix sizes $N$.
Right panel: dependence of $\gamma_1$ on $(\ln N)^{-1}$ (red line with 
crosses). 
The green line corresponds to the fit 
$\gamma_1(N) = \gamma_1(\infty) + \Delta\gamma/\ln N$ for 
the range $10^5\le N\le 10^9$ (i.e. $(\ln N)^{-1}<0.09$) 
with $\gamma_1(\infty)=1.020\pm 0.006$ and $\Delta\gamma=7.14\pm 0.09$.
}
\label{fig9}
\end{indented}
\end{figure}

To study the evolution of the eigenvalue spectrum with $N$ it is 
actually convenient to introduce the variable $\gamma_j = - 2 \ln |\lambda_j|$.
The dependence on $\gamma_j$ on the index $j$ is shown in left panel of 
Fig.~\ref{fig9}. 
It appears that the $\gamma$-spectra for different values 
of $N$ fall roughly on the 
same curve except for the last one or two values of each spectrum. 
This universal curve can be roughly 
approximated by a piecewise linear function with two 
slopes $\approx 4/3$ for $0\le j\le 6$ and $\approx 1/7$ for $6\le j\le 28$. 

We note that the convergence of the first nonzero $\gamma_1$ is compatible 
with the law 
$\gamma_1(N) \approx \gamma_1(\infty) + \Delta\gamma/\ln N$ with 
$\gamma_1(\infty)=1.020\pm 0.006$ and $\Delta\gamma=7.14\pm 0.09$ 
obtained from a fit in the range $10^5\le N\le 10^9$. This fit is actually 
very accurate as can be seen from the small error of $\gamma_1(\infty)$ 
and the right panel of Fig.~\ref{fig9}. 
Once more, such a dependence indicates a very slow logarithmic
convergence with the system size $N$.

\begin{figure} 
\begin{indented}\item[]
\begin{center}
\includegraphics[width=0.80\textwidth]{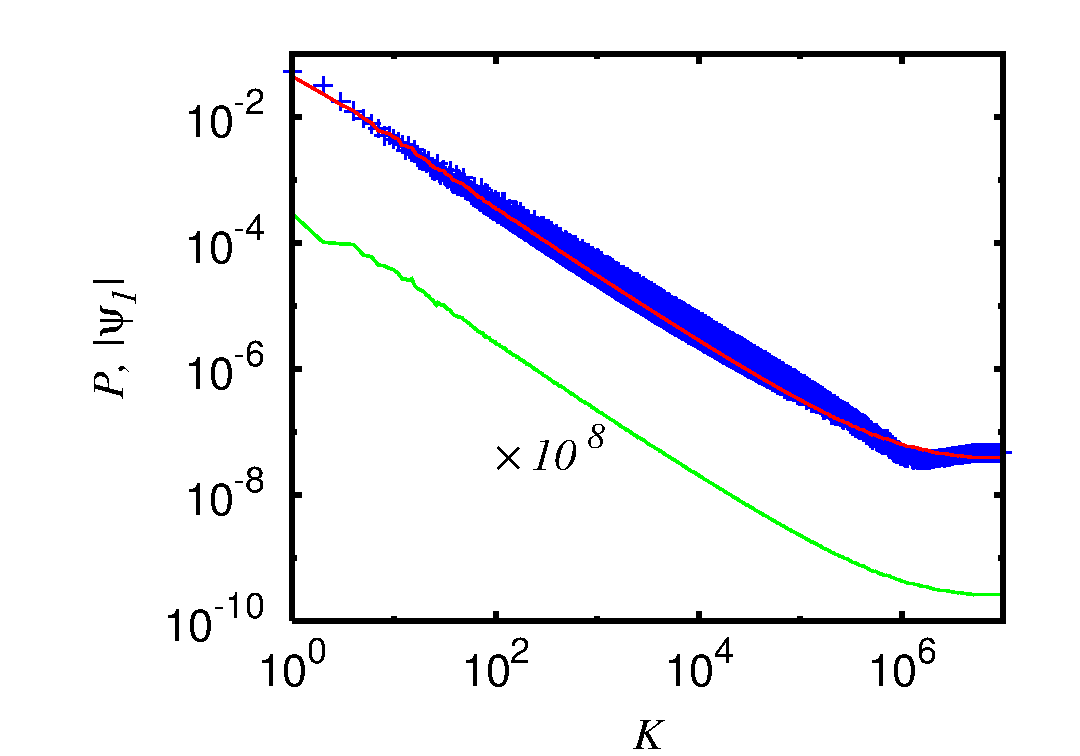}
\end{center}
\vglue -0.2cm
\caption{Dependence of the PageRank vector $P$ (red curve)
and the eigenvector $|\psi_1|$ (blue crosses) on
PageRank index K for $N=10^7$. Here the eigenvalue is
$\lambda_1=-0.28422 + i\, 0.38726$ 
($|\lambda_1|=0.48037$, $\gamma_1=1.4663$;
and the corresponding $\psi_1$ is normalized by the condition
$\sum_n|\psi_1(n)|=1$); green curve shows the difference
$|\Delta P|$ between the numerically computed PageRank $P$ (red curve)
and semi-analytical computation of PageRank;
for clarity $|\Delta P|$ is multiplied by a factor $10^8$. 
}
\label{fig10}
\end{indented}
\end{figure}

In Fig.~\ref{fig10}. we show the amplitude 
$|\psi_1|$ of 
the second eigenvector $\psi_1$ at $\lambda_1=-0.28422 + i\, 0.38726$ 
for $N=10^7$ versus the $K$ index. Despite some fluctuations this eigenvector 
seems to be close to the PageRank as far as the overall 
distribution of very large and 
small values is concerned. This behavior does not come as a surprise 
in view of the expansion [see Eq. (\ref{eq4})]~: 
\begin{equation}
\label{eq_vec2}
\psi_1\propto \sum_{j=0}^{l-1}\lambda_1^{-j-1}\,v^{(j)}\quad.
\end{equation}
In principle the fact that $|\lambda_1|$ is well below 1 indicates that 
the contributions of $v^{(j)}$ for larger values of $j$ increase. However, 
as we will discuss in the next section, the overall size of $v^{(j)}$ 
decays with increasing $j$ 
much faster than the increase by the factor $\lambda_1^{-j-1}$ and 
therefore mainly the first few terms of this sum contribute to $\psi_1$ in a 
similar way as for the PageRank (see Section \ref{sec5}). 

Finally in Fig.~\ref{fig10}, also 
the numerical difference of the 
PageRank determined by the standard power method and the semi-analytical 
expression (\ref{eq_PR1}) is shown. The relative difference is $\sim 10^{-10}$ 
for the full range of $K$ thus numerically confirming the accuracy of 
Eq. (\ref{eq_PR1}). 

\subsection{Numerical problems due to Jordan blocks}

The question arises why the Arnoldi Method for both initial vectors, 
random and uniform, (and also direct numerical 
diagonalization for small matrix sizes $N\le 10^4$) fail to confirm the 
analytical result that there are only $l=[\log_2(N)]$ non-zero eigenvalues 
$\lambda\neq 0$ of $S$. The reason is that the big subspace of 
dimension $N-l$ associated to the eigenvalue $\lambda=0$ with a lot of 
large Jordan blocks is numerically very problematic. This effect for 
such a {\em defective eigenvalue} is well known in the theory of numerical 
diagonalization methods \cite{arnoldibook}. To understand this a bit clearer 
consider a ``perturbed'' Jordan block of size $D$:
\begin{equation}
\label{eq7}
\left(\begin{array}{ccccc}
0 & 1 & \cdots & 0 & 0 \\
0 & 0 & \cdots & 0 & 0 \\
\vdots & \vdots & \ddots & \vdots & \vdots \\
0 & 0 & \cdots & 0 & 1 \\
\varepsilon & 0 & \cdots & 0 & 0 \\
\end{array}\right)
\end{equation}
which has a characteristic polynomial $\lambda^D-(-1)^{D}\varepsilon$ and 
therefore complex eigenvalues that scale as 
$|\lambda|\sim \varepsilon^{1/D}$ as 
a function of the perturbation $\varepsilon$ while for $\varepsilon=0$ 
we have $\lambda=0$ with multiplicity $D$. Therefore a value of 
$\varepsilon\sim 10^{-15}$ due to numerical rounding errors 
may still produce strong numerical errors in the eigenvalues if $D$ 
is sufficiently large. In our case Fig.~\ref{fig8} shows that 
the eigenvalues obtained by the Arnoldi method are accurate for 
$|\lambda|\ge 10^{-2}$. 

As can be seen in Fig.~\ref{fig8} there is also a difference in 
quality between the two initial vectors chosen for the Arnoldi Method. 
Using a random initial vector the Arnoldi method produces some wrong 
isolated eigenvalues in the intermediate regime $0.01\le |\lambda|\le 0.02$ 
and in the case $N=10^7$ some of the semi-analytical eigenvalues in the same 
regime are not accurately found. 
However, for the uniform initial vector the Arnoldi method produces 
rather accurate eigenvalues even for $|\lambda|\approx 0.005$. The reason is 
that the uniform initial vector corresponds (up to normalization) to the 
vector $v=e/N$. In view of Eq. (\ref{eq_represent1}) the Arnoldi method 
generates, at least in theory, exactly the $l$-dimensional subspace 
spanned by the vectors $v^{(j)}$ and should exactly break off at $n_A=l$ 
with a vanishing coupling matrix element from the subspace to the remaining 
space. However, due to numerical rounding errors and the fact that 
the vectors $v^{(j)}$ are badly conditioned, i.e. mathematically there are 
linearly independent but numerically nearly linearly dependent, the coupling 
matrix element is of order $10^{-3}$ (for $N=10^7$). As a consequence the 
Arnoldi method continues to generate new vectors producing a cloud of 
``artificial'' eigenvalues inside a circle or radius $\sim 0.005$. These 
eigenvalues are generated by the above explained mechanism of perturbed 
Jordan blocks. 

The Arnoldi method with a random initial vector produces a similar, slightly 
larger cloud, of such artificial eigenvalues but here, even without any 
numerical rounding errors, the method should not break off due to a bad choice 
of the initial vector and actually it even produces some ``bad'' eigenvalues 
outside the Jordan block generated cloud. 

We mention that it is possible to improve the numerical behavior of the 
Arnoldi method with uniform initial vector by the following ``tricks'': 
first we chose a different matrix representation of $S$ where the first basis 
vector (associated to the number ``1'') is replaced by the uniform vector 
$e$ and second where the 
scalar product used for the Gram-Schmidt orthogonalization is 
modified with stronger weights $\sim n^2$ for the larger components. This 
modified Arnoldi method produces a very small coupling matrix element 
$\sim 10^{-10}$ (for $N=10^7$) at $n_A=l$ and numerically very accurate 
eigenvalues 
(up to $10^{-10}$) for {\em all} $l$ non-vanishing eigenvalues. If we 
force to continue the Arnoldi iterations ($n_A\gg l$) we obtain again 
a Jordan block generated cloud of eigenvalues but whose size is 
considerably reduced as compared to both original variants of the method.

\section{Self-consistent determination of PageRank and analytic approximation}
\label{sec5}

The eigenvalue equation of the PageRank: $P=C\,v+S_0\,P$ with 
$C=d^{\,T}\,P$ [see Eq. (\ref{eq2})] can be interpreted as a 
self-consistent equation for $P$ defining a very effective iterative method to 
determine $P$ in a few number of iterations. 
Let us define the following iteration procedure~:
\begin{equation}
\label{eq_iter}
P^{(0)}=0\quad,\quad P^{(j+1)}=C\,v+S_0\,P^{(j)}\quad,
\quad j=0,\,1,\,2,\,\ldots\ .
\end{equation}
In principle the constant $C=d^{\,T}\,P$ is only obtained once the exact 
PageRank is known. Therefore in a practical application of this iteration, 
one first chooses some arbitrary non-vanishing value for $C$ and normalizes 
the PageRank once the procedure has converged. 
However, for reasons of notations we chose to keep the 
value $C=d^{\,T}\,P$ in Eq. (\ref{eq_iter}) from the very beginning. 

We note that the iteration (\ref{eq_iter}) can formally be solved by the sum~:
\begin{equation}
\label{eq_itersum}
P^{(j)}=C\,\sum_{i=0}^{j-1} S_0^i\,v
=C\,\sum_{i=0}^{j-1} v^{(i)}\ .
\end{equation}
Since $S_0^l=0$ for $l=[\log_2(N)]$ the iteration not only converges but it 
actually provides the exact PageRank $P=P^{(l)}$ 
after a finite number of iterations when $j=l$ and in which case 
Eq. (\ref{eq_itersum}) coincides with our previous result (\ref{eq_PR1}). 

\begin{figure} 
\begin{indented}\item[]
\begin{center}
\includegraphics[width=0.80\textwidth]{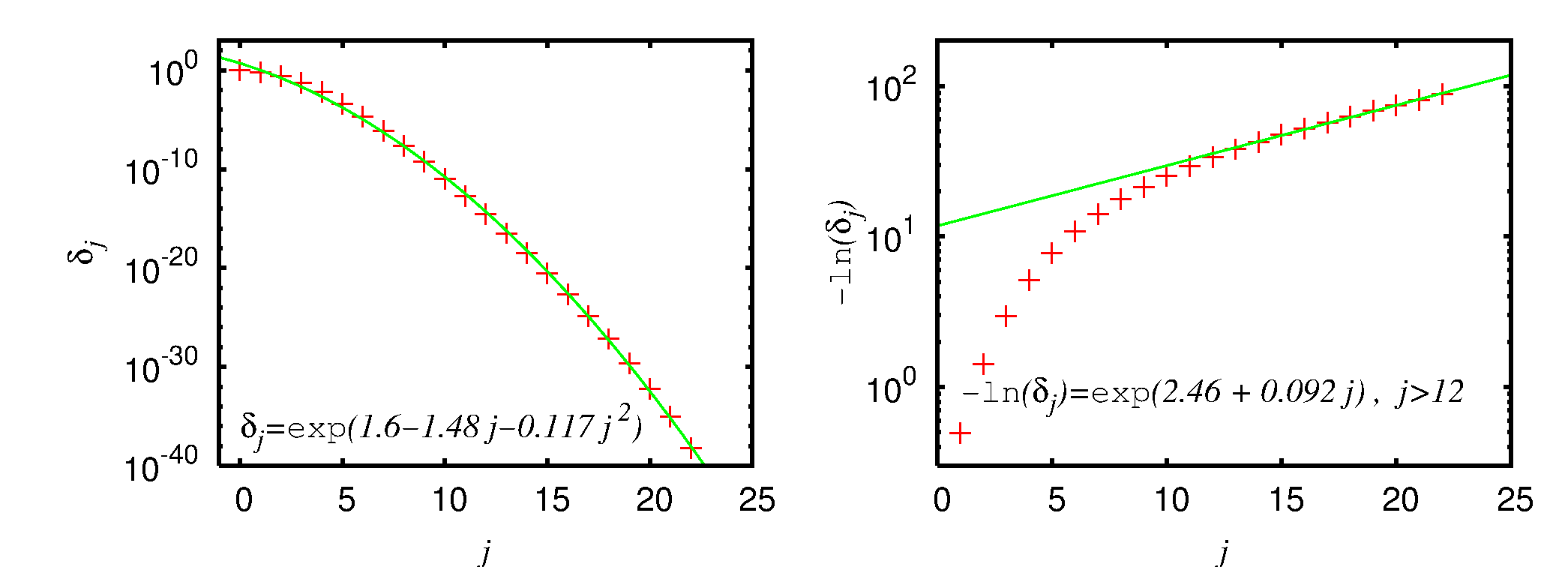}
\end{center}
\vglue -0.2cm
\caption{Decay of the quantity 
$\delta_j=\|P^{(j)}-P\|_1$ 
representing the error of the approximate PageRank $P^{(j)}$ after $j$ 
iterations of Eq. (\ref{eq_iter}) (for $N=10^7$). 
The left panel shows $\delta_j$ versus $j$ and the green line is obtained 
from the fit: $\ln(\delta_j)=a_3-b_3\,j-c_3\,j^2$ 
with $a_3=1.6\pm 0.4$, $b_3=1.48\pm 0.08$ and $b_3=0.117\pm 0.004$. 
The right panel shows $-\ln(\delta_j)$ 
versus $j$ and the green line is obtained 
from the fit: 
$\ln[-\ln(\delta_j)]=a_4+b_4\,j$ for $j>12$ with 
$a_4=2.46\pm 0.03$ and $b_4=0.092\pm 0.002$. 
Note that both panels use a logarithmic representation for the 
vertical axis. }
\label{fig11}
\end{indented}
\end{figure}

We mention that the power method, where one successively multiplies the matrix 
$S=v\,d^{\,T}+S_0$ to an initial (normalized) vector is somewhat similar 
to (\ref{eq_iter}) but with a very crucial difference. In the power method 
the constant $C$ is updated at each iteration according to 
$C^{(j)}=d^{\,T}\,P^{(j)}$ and here the initial vector must be different 
from $0$. We remind that the power method converges exponentially 
with an error 
$\sim |\lambda_1|^j$ where $\lambda_1$ being the second eigenvalue of $S$ with 
$|\lambda_1|\approx 0.5$ for $N=10^9$ and an extrapolated value 
$|\lambda_1|\approx 0.6$ in the limit $N\to\infty$. As can be seen 
in Fig.~\ref{fig11}, the iteration (\ref{eq_iter}) 
actually converges much faster than $|\lambda_1|^j$ which is simply due to 
fixing the constant $C$ from the beginning and not updating it with the 
iterations. 

The norm $\delta_j=\|P^{(j)}-P\|_1$ of the error vector after $j$ iterations 
decays much faster than exponentially with $j$ as it is shown in Fig.~\ref{fig11}.
For $N=10^7$ one can quite well approximate the error norm by 
the fit~: $\delta_j\approx \exp(1.6-1.48\,j-0.117\,j^2)$ representing 
a quadratic function in the exponential. Furthermore, for $j$ close to $l$ 
we have the approximate ratio $\delta_{j}/\delta_{j-1}\approx 10^{-2}$ 
and not $0.5$-$0.6$ as the power method would imply. For $j>12$ one can 
actually identify a regime of superconvergence where the logarithm of 
the error behaves exponentially~: $-\ln(\delta_j)\approx \exp(2.46+0.092\,j)$
but the parameter range for $j$ is too small to decide if 
there is really superconvergence. However, both fits clearly indicate 
that the convergence is considerably faster than exponential.

\begin{figure} 
\begin{indented}\item[]
\begin{center}
\includegraphics[width=0.80\textwidth]{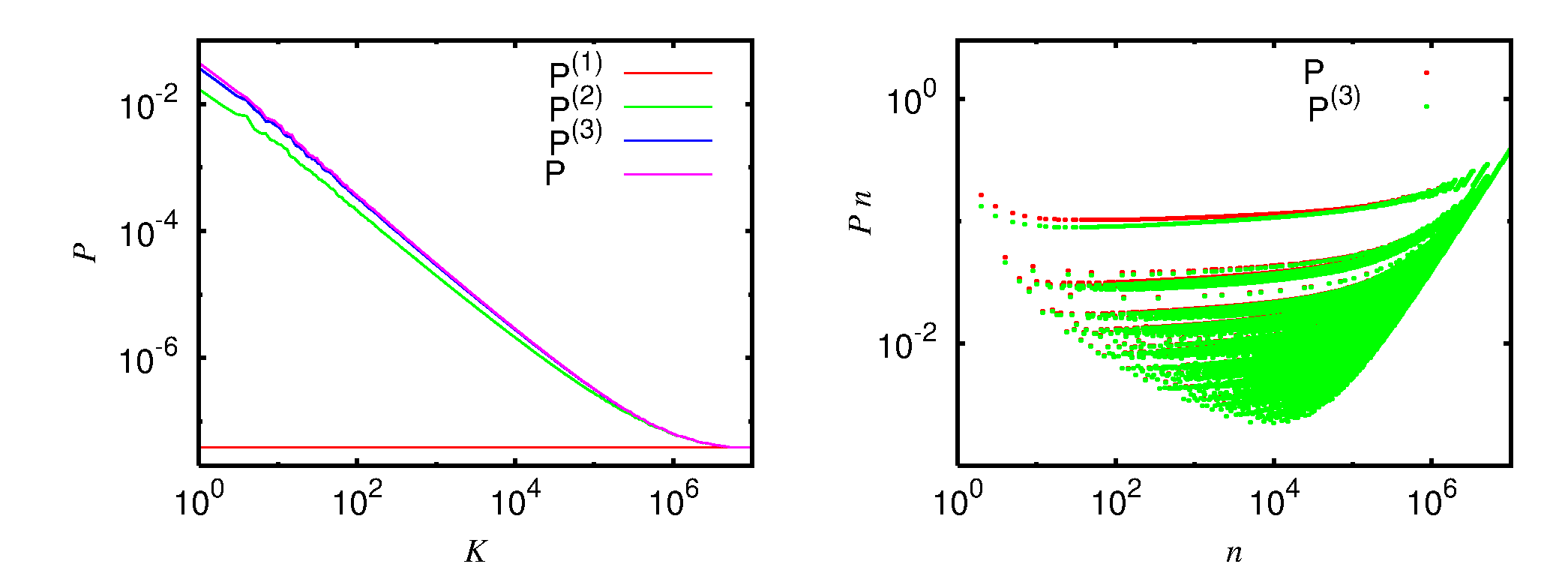}
\end{center}
\vglue -0.2cm
\caption{Left panel: Comparison of the first three PageRank approximations 
$P^{(j)}$ for $j=1,\,2,\,3$ obtained from Eq. (\ref{eq_iter}) 
and the exact PageRank $P$ versus 
the PageRank index $K$. 
Right panel: Comparison of the dependence of the rescaled probabilities 
$nP$ and $nP^{(3)}$ on $n$.
Both panels correspond to the case $N=10^7$. 
}
\label{fig12}
\end{indented}
\end{figure}

As a consequence of the very rapid convergence  dependent on the 
required precision, it is sufficient to apply the iteration (\ref{eq_iter}) 
only a few number of times $j\ll l$ to obtain a reasonable approximation. 
For example, Fig.~\ref{fig12} shows for $N=10^7$ 
that on a logarithmic scale 
$P^{(3)}$ and $P$ are already very close. 

This allows to obtain a very simple analytical approximation of the 
PageRank~: $P\approx P^{(3)}=v^{(0)}+v^{(1)}+v^{(2)}$. For this let us 
rewrite the recursion  $v^{(j+1)}=S_0\,v^{(j)}$ in a different way~:
\begin{equation}
\label{eq_diff1}
v^{(j+1)}_n=\sum_{m=2}^{[N/n]}\,\frac{M(mn,m)}{Q(mn)}\,v^{(j)}_{mn}
\quad{\rm if}\quad n\ge 2\quad{\rm and}\quad v^{(j+1)}_1=0\quad,
\end{equation}
where for given two integers $n$ and $m>1$ the multiplicity $M(n,m)$ is 
the largest integer such that $m^{M(n,m)}$ is a divisor of $n$ and 
$Q(n)=\sum_{m=2}^{n-1}M(n,m)$ is the number of divisors of $n$ (different 
from $1$ and $n$ itself) counting 
divisors several times according to their multiplicity. 
The appearance of the multiplicity $M(mn,n)$ in (\ref{eq_diff1})
is not very convenient for numerical evaluations. Either one recalculates 
the multicity at each use or one sacrifices a big amount of memory 
to store them. It is actually possible to 
rewrite Eq. (\ref{eq_diff1}) 
in a way that the multiplicities no longer appear explicitely. 
For this we note that the case $M(mn,n)\ge 2$ implies only to those 
values of $m$ such that $n$ is a divisor of $m$ implying $m=\tilde mn$ and 
$mn=\tilde mn^2$. This produces a second sum where one uses the 
multiples of $n^2$ and in a similar way a further sum with multiples of 
$n^3$ for the cases $M(mn,n)\ge 3$ and so on. For $n\ge 2$, we may 
therefore rewrite Eq. (\ref{eq_diff1}) in the following 
equivalent expression~:
\begin{equation}
\label{eq_diff2}
v^{(j+1)}_n=\sum_{m=2}^{[N/n]}\,\frac{1}{Q(mn)}\,v^{(j)}_{mn}
+\sum_{\nu\ge 2}^{n^\nu\le N} \sum_{m=1}^{\ [N/n^\nu]}
\,\frac{1}{Q(mn^\nu)}\,v^{(j)}_{mn^\nu}
\end{equation}
where each term in the sum of $\nu$ takes into account for the contributions 
with $M(mn,m)=\nu$. Note that the extra sums start at $m=1$ since 
$n\ge 2$ and therefore $mn^\nu> n$ even for $m=1$. 
The above PageRank iteration (\ref{eq_iter}) can be written in a similar way 
(see below) but for practical purposes, numerical or analytical, it is 
actually more convenient to use the recurrence for the vectors $v^{(j)}$ and 
to add them to obtain the PageRank according to Eq. (\ref{eq_itersum}). 

Both Eqs. (\ref{eq_diff1}) and (\ref{eq_diff2}) are also very efficient for 
a numerical evaluation, especially in terms of memory usage, 
since the matrix $S_0$ is 
represented by ``only'' $N$ integer values $Q(n)$, $n=1,\,\ldots,\,N$ which 
is much less than the number ($\sim N\ln N$) of non-zero 
double-precision matrix elements of $S_0$ 
(even completely taking into account the sparse structure of $S_0$). When 
using Eq. (\ref{eq_diff1}) one can recalculate at each time the 
multiplicities $M(n,m)$ which is not very expensive. However, it turns 
out that the additional sums in Eq. (\ref{eq_diff2}) are slightly more 
effective than this recalculation. 
Furthermore, for the iteration of $v^{(j)}$ the number of non-vanishing 
elements is reduced by a factor of two at each iteration. As a consequence 
we may replace in Eqs. (\ref{eq_diff1}) and (\ref{eq_diff2}) $N$ by 
$[N\,2^{-j}]$ and thus considerably reduce the computation time. We note 
that the direct iteration of $P^{(j)}$ instead $v^{(j)}$ does not have this 
advantage. Actually, in terms of numerical computation time the approximation 
to stop after few iterations is not very important since in any case the 
higher order corrections require less computation time. Using the iteration 
(\ref{eq_diff2}), we have been able to determine numerically the 
vectors $v^{(j)}$ and therefore the PageRank, the coefficients $c_j$ and the 
resulting $l=[\log_2 N]$ non-zero eigenvalues of $S$ for system sizes up to $N=10^9$. 

In addition,  Eq. (\ref{eq_diff1}) allows also for some analytical approximate 
evaluation of the first vectors. The initial vector is $v^{(0)}_n=1/N$. 
Let us try to evaluate the next two vectors $v^{(1)}_n$ and $v^{(2)}_n$ for 
the most important case where $n$ is a prime number $p$. 
Furthermore, in the sum (\ref{eq_diff1}) the most important contributions 
arise for $m$ also being a prime number 
$q$ such that $Q(qp)=2$ and $M(qp,p)=1$ (except for the case $q=p$ which we 
neglect) resulting in~:
\begin{equation}
\label{v1_approx1}
v^{(1)}_p\approx \sum_{q=2,{\rm\ prime}}^{[N/p]} \frac{1}{2N}
=\frac{1}{2N}\,\pi\left(\left[\frac{N}{p}\right]\right) 
\approx\frac{1}{2p(\ln N-\ln p)}
\end{equation}
where $\pi(n)\approx n/\ln(n)$ (for $n\gg 1$) is the number of prime numbers 
below $n$. However, these values of $v^{(1)}_n$ at prime numbers $n=p$ do not 
contribute in (\ref{eq_diff1}) for the next iteration 
$j=1$ when trying to determine 
$v^{(2)}$. To obtain the leading contributions in $v^{(2)}$ 
we need $v^{(1)}_n$ for $n=p_1\,p_2$ being a product of 
two prime numbers. In this case, we have $Q(q\,p_1\,p_2)=2^3-2=6$ if 
$q,\,p_1,\,p_2$ are three different prime numbers. Assuming $p_1\neq p_2$ 
and neglecting the 
complications from the few cases $q=p_1$ or $q=p_2$, we find that~:
\begin{equation}
\label{v1_approx2}
v^{(1)}_{p_1\,p_2}\approx 
\frac{1}{6N}\,\pi\left(\left[\frac{N}{p_1\,p_2}\right]\right) 
\approx\frac{1}{6p_1\,p_2\,(\ln N-\ln p_1-\ln p_2)}\ .
\end{equation}
For the case $n=p^2$, i.e. $p_1=p_2=p$, we have 
$Q(qp^2)=5$ (since $p$ has multiplicity 2) resulting in~:
\begin{equation}
\label{v1_approx3}
v^{(1)}_{p^2}\approx 
\frac{1}{5N}\,\pi\left(\left[\frac{N}{p^2}\right]\right) 
\approx\frac{1}{5p^2\,(\ln N-2\,\ln p)}\ .
\end{equation}
From (\ref{eq_diff1}) for $j=1$ and (\ref{v1_approx2}) we obtain~:
\begin{equation}
\label{v2_approx1}
v^{(2)}_p\approx \frac{1}{12N}\sum_{q=2,{\rm\ prime}}^{[N/(2p)]} \pi\left(\left[
\frac{N}{p\,q}\right]\right)\ .
\end{equation}
Here we have reduced the sum from $q\le [N/p]$ to $q\le [N/(2p)]$ since 
$\pi\left([N/(pq)]\right)$ is non zero only for $N/(pq)\ge 2$ and therefore 
$q\le N/(2p)$. Now, we replace the sum $\sum_q (\cdots)$ over the prime 
numbers by an integral 
$\int dq\,\pi'(q)\,(\cdots)$ where $\pi'(q)\approx 1/\ln(q)$ is the 
average density of prime numbers at $q$ resulting in~:
\begin{eqnarray}
\nonumber
v^{(2)}_p &\approx& \frac{1}{12N}\int_2^{N/(2p)} dq\,
\pi\left(\left[\frac{N}{p\,q}\right]\right)\,\pi'(q)\\
\nonumber
&\approx& \frac{1}{12p} \int_2^{N/(2p)} \frac{dq}{q}\, \frac{1}{
\Bigl(\ln(N/p)-\ln q\Bigr)\ln q}\\
\nonumber
&=& \frac{1}{12p} \int_{\ln 2}^{\ln\left(N/(2p)\right)} du\,
\frac{1}{\Bigl(\ln(N/p)-u\Bigr)\,u}\\
\label{v2_approx2}
&=& \frac{1}{6p\,\ln(N/p)}\left(
\ln \ln \left(\frac{N}{2p}\right)-\ln \ln 2\right)\ .
\end{eqnarray}
From (\ref{v1_approx1}) and (\ref{v2_approx2}) we obtain 
the PageRank approximation at integer values 
\begin{equation}
\label{page_approx}
P_p\approx P^{(3)}_p\approx C(\frac{1}{N}+v^{(1)}_p+v^{(2)}_p)
\approx \frac{C}{2p\ln N}\left(1-\ln\ln 2+\frac{\ln\ln N}{3}\right)
\end{equation}
where we have assumed that $N\gg p$ and replaced 
$\ln(N/p)=\ln N-\ln p\approx \ln N$ and $C$ is the same constant 
as used in (\ref{eq_iter}). 

The important point with this expression is 
that it is of the form $P_p\approx C_N/p$ where $C_N$ is a constant 
depending on $N$. In order to compare with our above results, especially 
in Fig.~\ref{fig2}, we have to replace $p$ by the $K$ index. Assuming 
that the $K$ index is dominated by the prime numbers we have 
$K=\pi(p)\approx p/\ln p$ implying $p\approx K\ln p\approx 
K\,\ln K$ thus providing the behavior $P(K)\approx C_N/(K\,\ln K)$ 
already conjectured above based on the numerical results. Concerning the 
numerical value of the constant $C_N$ we find that, at $N=10^7$, 
it is roughly one order 
of magnitude too small as compared to the numerical results. 

We remind that the considerations leading to the expression 
(\ref{page_approx}) are based on a lot of assumptions and quite crude 
approximations, especially the replacement of $\pi(n)\approx n/\ln(n)$, 
even if $n={\cal O}(1)$, and we have neglected a lot of contributions 
from numbers with more factors in their prime factor decomposition which 
are most likely responsible for the reduced numerical prefactor. 
Furthermore, the assumption that the PageRank is dominated by prime numbers 
is not completely exact since certain non-prime 
numbers with a small number of factors 
intermix with larger prime numbers in the PageRank, thus modifying 
the dependence of the prime numbers on the $K$ index  
from $p\approx K\,\ln(K)$ to $p\approx K\,(1.36+0.323\ln K)$ 
according to the fit in Fig.~\ref{fig6} for $N=10^7$. However, 
despite the approximations, we recover the leading parametric dependence 
of $P\sim 1/(K\,\ln K)$.

\begin{figure} 
\begin{indented}\item[]
\begin{center}
\includegraphics[width=0.80\textwidth]{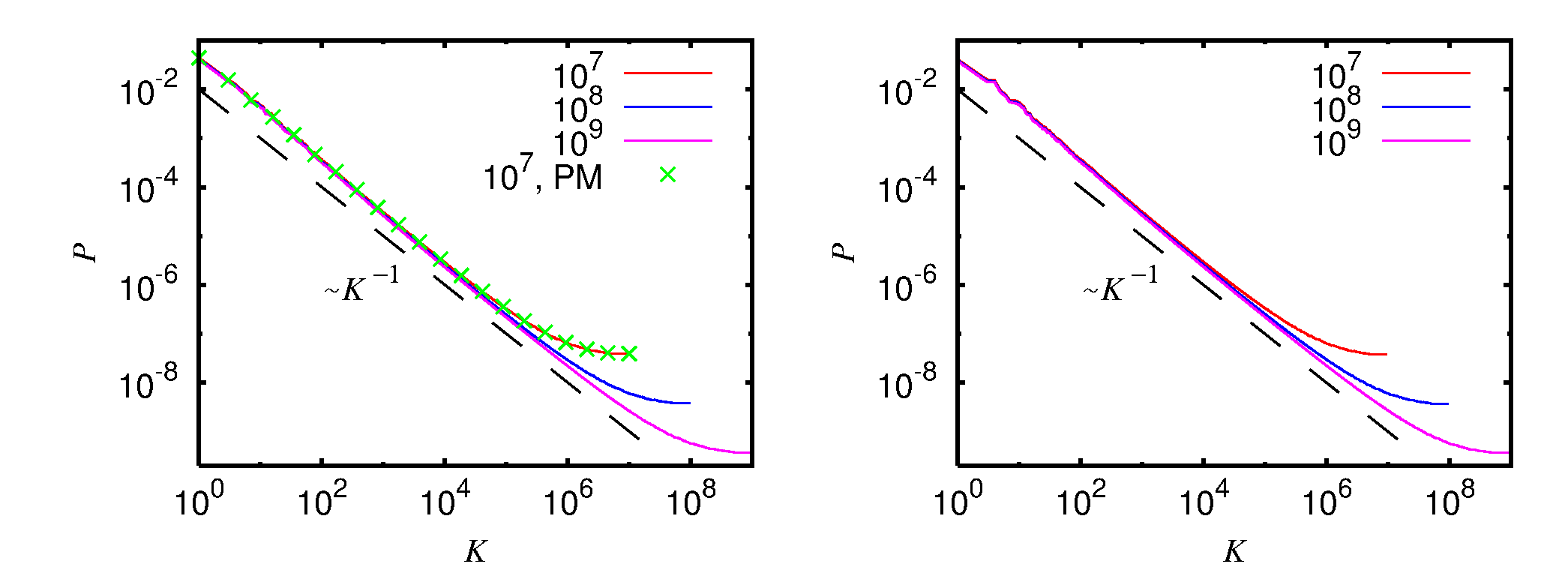}
\end{center}
\vglue -0.2cm
\caption{Left panel: The full lines correspond to the 
dependence of PageRank probability $P(K)$
on PageRank index $K$ for the matrix sizes 
$N=10^7$, $10^8$, $10^9$ with the PageRank evaluated from the 
expression (\ref{eq_PR1}) using the efficient numerical method 
based on Eq. (\ref{eq_diff2}). The green crosses correspond 
to the PageRank obtained by the power method (PM) for $N=10^7$;
the dashed straight line shows the Zipf law dependence $P \sim 1/K$. 
Right panel: same as in left panel (without data from the power method)
for a simplified model for the 
Google matrix of integers where all multiplicities $M(n,m)$ are 
replaced by 1, i.e. $n$ is to linked to its divisors $m$ only once 
even if $n$ can be divided several times by $m$. The PageRank was numerically 
evaluated by the same efficient method using Eqs. (\ref{eq_PR1}) and 
(\ref{eq_diff1}) with $M(n,m)=1$. 
}
\label{fig13}
\end{indented}
\end{figure}

\begin{figure} 
\begin{indented}\item[]
\begin{center}
\includegraphics[width=0.80\textwidth]{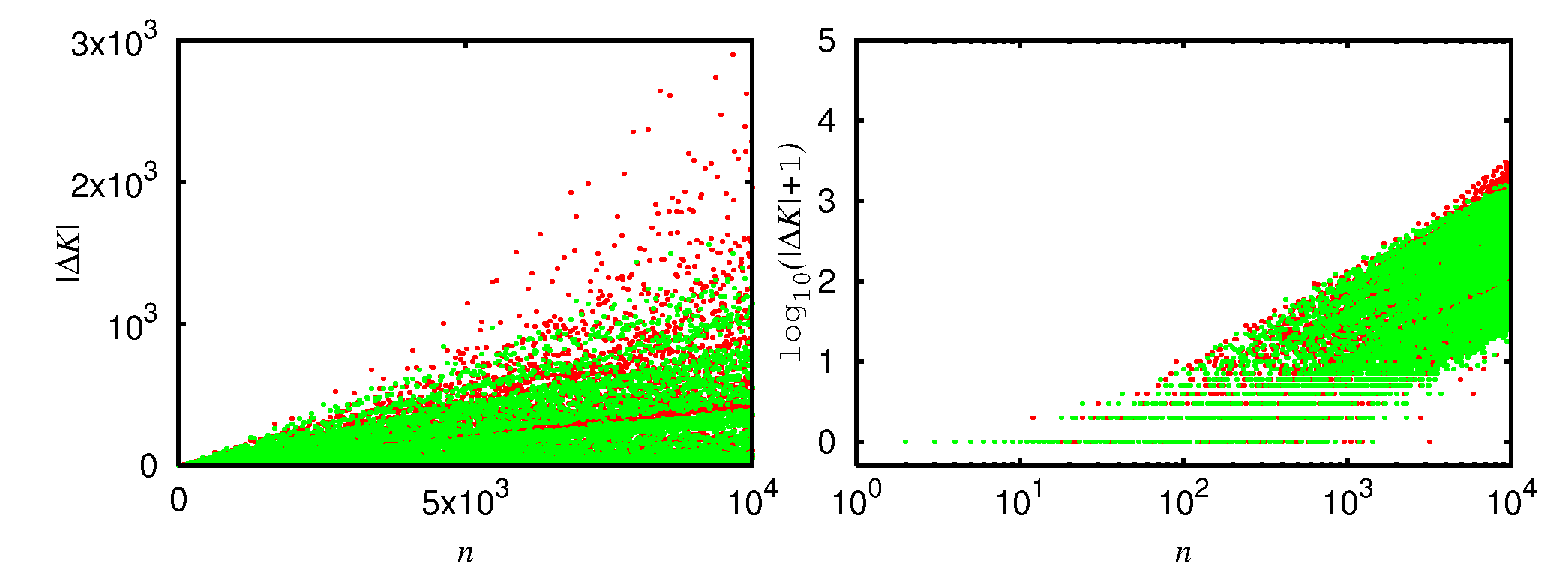}
\end{center}
\vglue -0.2cm
\caption{Dependence of $|\Delta K| = |K_n(N_2)-K_n(N_1)|$
on integer $n$ for matrix sizes $N_1=10^8, N_2=10^9$
(green points) and
$N_1=10^7, N_2=10^8$ (red points).
Left and right panels show the same data
in normal and log-log scales.
Note the strongly reduced vertical scale of the left panel 
as compared to the left panel of Fig.~\ref{fig7}. The vertical scale 
of the right panel was not reduced allowing a direct comparison with 
the right panel of Fig.~\ref{fig7}. The data was obtained by the same 
efficient numerical method as in the left panel of Fig.~\ref{fig13}. 
}
\label{fig14}
\end{indented}
\end{figure}

The PageRank dependence $P(K)$ obtained from the 
expression (\ref{eq_PR1}) using the efficient numerical method 
based on Eq. (\ref{eq_diff2}) is shown in Fig.~\ref{fig13} (left panel)
for $N=10^7, 10^8, 10^9$. For $N=10^7$ these data agree with the
computation result by the Arnoldi power method with the numerical accuracy of 
the order of $10^{-10}$ (see also Fig.~\ref{fig10}). This confirms 
the efficiency of our semi-analytical computation of the  PageRank.

We note that it may be useful to consider a simplified model
for the Google matrix of integers
when multiplicity of all divisors is taken to be unity.
The numerical fit of data shows that in this case the number of links
scales as $N_\ell=N\,(a_\ell+b_\ell \ln N)$ with
$a_\ell=-1.838 \pm  0.002$, $b_\ell=0.999 \pm  0.0002$.
For this model we have the same expression
 (\ref{eq_diff1}) but with the replacements 
$M(nm,m)\to 1$ and $Q(n)\to Q^*(n)$ 
where $Q^*(n)$ is the number of divisors of the integer $n$ excluding 1 
and $n$ itself without multiplicities, e.~g. 
$Q^*(2)=0,\, Q^*(3)=0,\, Q^*(4)=1,\, \ldots$. Note that this 
quantity is given by 
the expression $Q^*(n)=\left(\prod_j (\mu_j+1)\right)-2$ where $\mu_j$ 
are the exponents in the prime factor decomposition of 
$n=\prod_j p_j^{\mu_j}$. 

The dependence of the PageRank on $K$ for the simplified model is shown in 
the right panel of Fig.~\ref{fig13}. It shows practically the same behavior 
as in the
main model shown in the left panel. In this case the analytical expression 
for the PageRank $P$, obtained from the first three terms, has 
a very simple form
\begin{eqnarray}
\label{simmodel}
P_n\approx P^{(3)}_n = \sigma_N \Biggl(&1& + \sum_{m_1=1}^{[N/n]} 
\frac{1}{Q^*(m_1 n)} \\
\nonumber
&&+ \sum_{m_1=2}^{[N/n]} \sum_{m_2=2}^{\ [N/(nm_1)]} 
\frac{1}{Q^*(m_1 n)}\frac{1}{Q^*(m_2 m_1 n)} \Biggr)
\end{eqnarray}
where $N$ is the matrix size and $\sigma_N$
is the global normalization constant  determined by the condition
$\sum_{n=1}^{n=N} P_n =1$. This simple formula gives a good description of the
PageRank behavior shown in the right panel of Fig.~\ref{fig13}.
Indeed, the direct count shows that the ratio $R_{ms}$ of the total 
number of links $N_\ell$ for both models 
(counted with or without multiplicities) approaches to unity
for large matrix sizes. For example, we have $R_{ms}=1.184$ ($N=1000$),
$1.102$ ($10^5$), $1.070$ ($10^7$), $1.052$ ($10^9$). Thus we think that
in the limit of large $N$ both models converge to the same type of behavior.
It is possible that the simplified model may be more suitable for further
analytical analysis. However, in this work we present data for the simplified model
only in the right panel of Fig.~\ref{fig13}.

Using the PageRank data obtained by the self-consistent approach
for large $N=10^7, 10^8, 10^9$ we can analyze the convergence of 
the PageRank order $K_n$ at larger sizes compared to those used in 
Fig.~\ref{fig7}. These new results for variation of $|\Delta K|$
are presented in Fig.~\ref{fig14}. They show 
that the variation $|\Delta K|$ decreases with the increase of
$N$ from $10^7$ up to $10^9$ even if the process is slow.
A direct comparison  shows that the first deviation in the order $K_n$
appears at $K=K_s=13$ (comparing $N=10^6$ vs. $10^7$),
$K_s=27$ ($10^7$ vs. $10^8$), $K_s=30$ ($10^8$ vs. $10^9$).
We find that the stable range interval $K_s$
grows with $N$ but this growth seems logarithmic
like with $K_s \sim \ln N$. Such a growth seems to be natural in the
view of logarithmic convergence of the second eigenvalue $\lambda_1$ 
discussed above
and all logarithmic factors appearing in the density of primes.
We also note that the value of $K_s$ is significantly smaller than the 
value of $K_d$ at which the first degenerate flat plateau appears in 
the PageRank $P(K)$ and hence these degeneracies do not affect the order of 
the first $K_s$ integers.

On the basis of the obtained results we conclude that for our 
maximal matrix size $N=10^9$ we have convergence of the 
first 32 values of $K_n$. These numbers $n$,
corresponding to the values of $K=1,\,2,\,\ldots,\,32$, are
$n=2$, $3$, $5$, $7$, $4$, $11$, $13$, $17$, $6$, $19$, $9$, $23$, $29$, $8$, 
$31$, $10$, $37$, $41$, $43$, $14$, $47$, $15$, $53$, $59$, 
$61$, $25$, $67$, $12$, $71$, $73$, $22$, $21$. There are about 30\%
of non-primes among these values. We mention that the positions 
of the first non-primes $4,\,6,\,9$ can be already obtained 
from the first order approximations of $v^{(1)}$ discussed above. 
According to (\ref{v1_approx1}) the relative weight of a prime number 
in first order is $1/(2p)$. For the two square numbers $4$ and $9$ 
the weight is according to (\ref{v1_approx3}) either 
$1/(5\times 4)=1/(2\times 10)$ or $1/(5\times 9)=1/(2\times 22.5)$ 
explaining that $4$ is between the primes $7$ and $11$ and that $9$ 
is between $19$ and $23$. For the product $6=2\times 3$ we have 
according to (\ref{v1_approx2}) the weight 
$1/(6\times 6)=1/(2\times 18)$ implying that $6$ is between 17 and 19. 
However, this simple argument does not work for other numbers, for 
example for $10$ (or $14$) it would imply an incorrect position 
between $29$ and $31$ ($41$ and $43$). 
We mention that more numerical data
are available at the web page \cite{webpage}. 

For the simplified model we find at $N=10^9$ for the first
values $K=1,\,2,\,\ldots,\,32$ a slightly different order of integers
$n=2$, $3$, $5$, $4$, $7$, $11$, $13$, $17$,
$9$, $6$, $19$, $8$, $23$, $29$, $31$, $10$, $37$, $41$, $43$, $14$, 
$47$, $15$, $53$, $25$, $59$, $16$, $61$,
$12$, $67$, $71$, $22$, $21$. Here the absence of multiplicities 
increases the weight for square numbers of primes to $1/(4p^2)$ 
implying that these numbers are slightly advanced in the $K$ order as 
compared to our main model. 
The modified weight for $9$ is $1/(2\times 18)$ coherent with 
new position between $17$ and $19$ (with 
$6$ having the same first order weight as $9$ and also being between $17$ 
and $19$). For $4$ the 
weight is increased from $1/(2\times 10)$ to $1/(2\times 8)$. However, 
this increase is not sufficient to explain the new position of $4$ between $5$ 
and $7$.

One might mention as a curiosity a special ``prime integer network model''
where a non-prime 
number $n$ is only linked to its prime factors (and not to all of its 
divisors). 
In this case the matrix $S_0$ is strongly simplified such that $S_0^2=0$, i.e. 
$l=2$ being independent of the system size and hence 
there are only two non-vanishing 
eigenvalues of the Google matrix which are $\lambda_0=1$ and 
$\lambda_1=c_0-1\approx -1+1/\ln N$ where 
$c_0=(\pi(N)+1)/N\approx 1/\ln N$ is the 
ratio of the number of primes and unity to $N$. This is simply seen from 
the definition of $c_j$ in Eq. (\ref{eq5}) and the trace 
$c_0=\lambda_0+\lambda_1$ of the matrix (\ref{eq_represent2}) which 
is of size $2\times 2$ for this case. According to (\ref{eq4}) the 
PageRank $P$ 
and the second eigenvector $\psi_1$ are given by $P\propto e+v^{(1)}$ 
and $\psi_1 \propto e-v^{(1)}/(1-1/\ln N)$ where $e$ is the vector with 
all components equal to unity and $v^{(1)}$ is a vector such that 
$v_n^{(1)}=0$ for non-prime numbers $n$ or $n=1$ and 
$v_n^{(1)}$ for prime numbers $n=p$ is given by an equation 
similar to Eq. (\ref{eq_diff1}) for $j=0$ with 
$v_{nm}^{(0)}$ being replaced by 
unity and multiplicities and number of divisors adapted for the prime 
integer network model. Here both versions, with or without multiplicities are 
possible. The eigenvalues do not depend on the version but the eigenvectors 
do. For both cases it is pretty obvious that the $K$ index gives exactly 
the sequence of prime numbers below $N$ in increasing order followed 
by a large 
degenerated plateau for the non-prime integer numbers. Note that 
here the second eigenvalue converges to $-1$ with a correction $1/\ln(N)$ 
for large $N$ thus closing the gap in $|\lambda|$ of the Google matrix.

\section{Discussion}

In this work we constructed the Google matrix of integers 
based on links between a given integer $n$ and its divisors.
The numerical analysis based on the Arnoldi method
allowed us to show that the PageRank $P(K_n)$ of this directed network
decays with PageRank index $K_n$ of an integer $n$
approximately as $P(K_n) \sim 1/(K_n \ln K_n)$
being similar to those of the Zipf law and those 
found for the WWW. However, the spectrum of the Google
matrix has a large gap appearing between the unit eigenvalue
and other eigenvalues while the spectrum of the Google matrix of
WWW usually has no gap. We developed an efficient 
semi-analytical method to compute the PageRank 
of integers which allowed us to determine the dependence
$P(K_n)$ up to matrix size of one billion. We show that
the dependence of PageRank on the integer number $n$ is
characterized by a series of branches corresponding to 
primes, semi-primes and numbers with a higher products of primes.
Our data show a logarithmic like convergence of PageRank order of integers
to a fixed order in the limit of matrix size going to infinity.

\ack
This work is supported in part by  the EC FET Open project 
``New tools and algorithms for directed network analysis''
(NADINE $No$ 288956).

\end{document}